\documentclass[]{mn2e}
\usepackage{epsfig,graphicx,lscape}

\title[Structure of Magellanic I/HVCs]{Large and small-scale structure of the Intermediate and High Velocity Clouds towards 
the LMC and SMC}

\author[J. Smoker et al.]
{J.~V. Smoker$^{1}$\thanks{email: jsmoker@eso.org \hspace*{0.3cm} 
Based on ESO programme IDs 078.C-0493(A) and 
171.D-0237(B).},
A.~J. Fox$^{2}$, F.~P. Keenan$^{3}$
\\
$^{1}$European Southern Observatory,
Alonso de Cordova 3107,
Casilla 19001,
Vitacura,
Santiago 19, Chile \\
$^{2}$
Space Telescope Science Institute, 3700 San Martin Drive,
Baltimore MD 21218, USA \\
$^{3}$Astrophysics Research Centre,
School of Mathematics and Physics,
Queen's University Belfast, \\
Belfast, BT7 1NN            \\
}

\date{Accepted
Received
in original form }

\def\LaTeX{L\kern-.36em\raise.3ex\hbox{a}\kern-.15em
T\kern-.1667em\lower.7ex\hbox{E}\kern-.125emX}

\pubyear{}

\begin{document}
\label{firstpage}
\maketitle

\begin{abstract}

We employ Ca\,{\sc ii} K and Na\,{\sc i} D interstellar absorption-line spectroscopy of early-type stars in
the Large and Small Magellanic Clouds to investigate the large- and small-scale structure in foreground 
Intermediate and High Velocity Clouds (I/HVCs). 
These data include FLAMES-GIRAFFE Ca\,{\sc ii} K observations of 403 
stars in four open clusters, plus FEROS or UVES spectra of 156 stars 
in the LMC and SMC. The FLAMES observations are amongst the most extensive 
probes to date of Ca\,{\sc ii} structures on $\sim$20 arcsec scales  
in Magellanic I/HVCs.

From the FLAMES data within a 0.5$^{\circ}$ field-of-view, the
Ca\,{\sc ii} K equivalent width in the I/HVC components towards three clusters 
varies by factors of $\ge$10. 
%
There are no detections of molecular gas in absorption at intermediate or high velocities, although 
molecular absorption is present at LMC and Galactic velocities towards some sightlines.
The sightlines show variations in EW exceeding a factor $\sim$7 in CH$^{+}$ towards NGC\,1761 over 
scales of less than 10 arcminutes. 

The FEROS/UVES data show Ca\,{\sc ii} K I/HVC absorption in $\sim$60 per cent of sightlines. 
No Na\,{\sc i} D is found at non-Magellanic HVC velocities aside from a tentative detection towards 
the star LHA 120-S 93. The range in the Ca\,{\sc ii}/Na\,{\sc i} ratio in I/HVCs is from  --0.45 to +1.5 dex, 
similar to previous measurements for I/HVCs. 
In ten sightlines we find Ca\,{\sc ii}/O\,{\sc i} ratios in I/HVC gas ranging from 0.2 to 1.5  
dex below the solar value, indicating either dust or ionisation effects.  
In nine sightlines I/HVC gas is detected in both H\,{\sc i} and Ca\,{\sc ii}, and shows
similar Ca\,{\sc ii}/H\,{\sc i} ratios to typical I/HVCs, and similar velocities, implying 
that in these sightlines the two elements form part of the same structure. 
\end{abstract}


\section{Introduction} \label{s_intro}

High velocity clouds (HVCs) were discovered over 50 years ago (Muller, Oort \& Raimond 1963). Originally observed 
in H\,{\sc i}, they consist of parcels of gas with velocities not compatible with Galactic rotation; 
in practice this means HVCs have $|v_{\rm LSR}|>$90--100\,km\,s$^{-1}$ if they lie at at high Galactic latitudes. 
Clouds with $\sim$40 $<$ $|v_{\rm LSR}|$ $<$ 90\,km\,s$^{-1}$ are referred to as Intermediate Velocity Clouds (IVCs). 
Recently it has become clear that at least some of the clouds lie within the halo of the Milky Way; 
for example Complex A (van Woerden et al. 1999), Complex WB (Thom et al. 2006), Complex C (Wakker et al. 2007, 
Thom et al. 2008), as well as the Cohen Stream, Complex GCP, and Complex g1 (Wakker et al. 2008). 

Although some HVCs may be large clouds outside of the Milky Way halo (Blitz et al. 1999), the 
failure to detect HVCs far from their host galaxies (Pisano et al. 2004; Westmeier, Br\"uns \& Kerp 2008) 
and the lack of stars in many HVCs (e.g. Simon \& Blitz 2002; Siegel et al. 2005; Hopp, Schulte-Ladbeck \& Kerp 2007) 
argues against this hypothysis. Likewise, comparison of the probability of detection of ionised HVCs in AGN and 
Galactic halo stellar sightlines by Lehner \& Howk (2011) indicate that HVCs at velocities between $\sim$90 and 
170 km\,s$^{-1}$ likely reside within the halo of the Milky Way and are not further away structures within the 
local group. These clouds may have formed by the accretion of primordial gas, interaction of the Milky Way 
with its neighbouring galaxies via ram pressure and tidal stripping, or the Galactic Fountain (see Bregman 
2004 for a review). Finally, Lehner \& Howk show that HVCs at extreme velocities ($>$170 km\,s$^{-1}$) lie beyond 
the Galactic Halo and are not discussed further here. 


I/HVCs are important to study as they {\em may} provide fuel for star formation in the Milky Way 
(see the review by Wakker \& van Woerden 1997; Fox et al. 2014), and provide information on 
close encounters and/or winds from the SMC and LMC (Olano 2008; Lehner, Staveley-Smith \& Howk 2009). 

In the present paper we employ FLAMES Ca\,{\sc ii} K archive spectra of four open clusters 
in the Large and Small Magellanic Clouds (LMC, SMC) to investigate the small-scale ($\sim$11 arcseconds to $\sim$25 arcminutes) 
structure of the I/HVCs in the lines-of-sight to these galaxies. We also use FEROS and UVES observations of 156 LMC and SMC 
stellar targets in Ca\,{\sc ii} K and  Na\,{\sc i} D to probe large- (degree-) scale variations in I/HVC column 
density. Observations of interstellar Ca and Na towards  the Magellanic Clouds was first performed by Blades (1980), with subsequent 
observations in different wavelength bands and theoretical work by Savage \& de Boer (1981), Songaila (1981), Songaila, Cowie \& York (1981), Songaila et al. (1986), 
Andreani \& Vidal-Madjar (1988), Blades et al. (1988a,b), 
Wayte (1990), Molaro et al. (1993), Welty et al. (1997, 1999), Richter et al. (1999), 
Bluhm et al. (2001), Staveley-Smith et al. (2003), Andr\'{e} et al. (2004), Olano (2008), 
Lehner et al. (2009) and Welty, Xue and Wong (2012). These and other references indicate that some  
I/HVCs contain molecular gas and dust, often show multi-component velocity structure, and have abundance 
patterns sometimes consistent with those of the Magellanic Clouds and sometimes with the Milky Way. 
The velocity fields of Magellanic HVCs have been interpreted as being caused by spiral structure in the LMC 
(Blades 1980), winds from the LMC (Olano et al. 2004), or interaction of LMC/SMC and Milky Way gas (Olano et al. 2008). 
Background sources in the Magellanic Clouds have also been used to study the absorption-line structure within the 
Milky Way (e.g. Andr\'{e} et al. 2004, Nasoudi-Shoar et al. 2010, van Loon et al. 2013; Smoker, Keenan \& Fox 2015). 

Our paper is laid out as follows. Section \ref{Sample} describes the sample of stars towards the Magellanic Clouds and a 
description of the data reduction for the optical spectra. In Sect. \ref{Results} we provide the main results, 
including the FLAMES, FEROS and UVES spectra towards the Magellanic system. Section \ref{Discussion} presents the discussion 
which covers the velocity dependence on RA of LMC I/HVCs, abundance ratios using optical and previous UV data, 
and large- and small-scale structure variations in Ca\,{\sc ii} K of the I/HVCs towards the LMC and SMC. Finally, 
Sect. \ref{Summary} gives a summary of the main findings. 

\section{The sample, observations and data reduction} 
\label{Sample}


\subsection{Archival FLAMES and FEROS data towards the LMC and SMC}

\begin{figure}
\epsfig{file=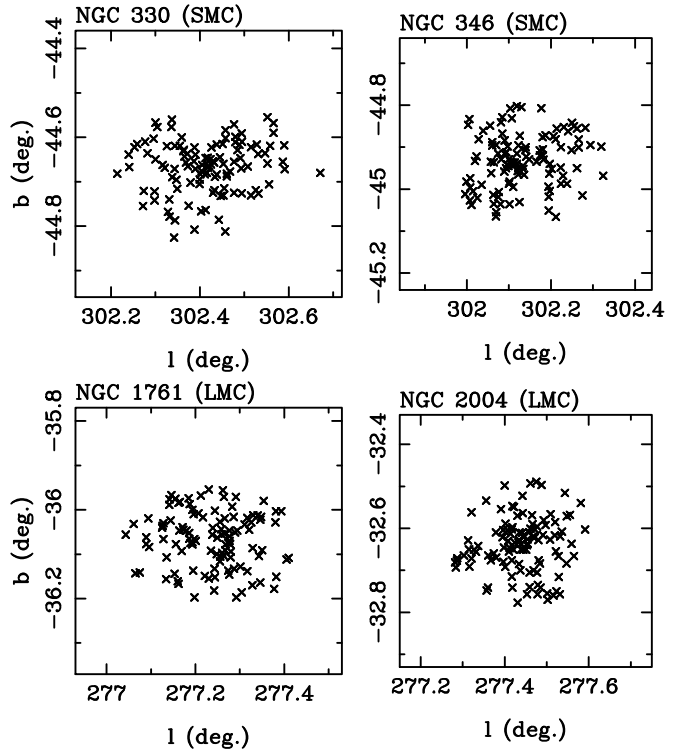}
\caption[]{Location of LMC and SMC stars observed with FLAMES-GIRAFFE.}
\label{f_MC_Clusters_lb_FLAMES}
\end{figure}

FLAMES\footnote{FLAMES (Pasquini et al. 2002) is a multi-object, intermediate and high
resolution spectrograph, mounted on the VLT/Unit Telescope 2 (Kueyen)
at Cerro Paranal, Chile, operated by ESO.} 
observations towards four open clusters in the LMC and SMC were retrieved from the ESO archive and 
are used to study the I/HVCs towards these galaxies and their variation on small scales. Table \ref{t_sample_FLAMES} presents 
the basic cluster data, and Fig. \ref{f_MC_Clusters_lb_FLAMES} shows
the locations of the stellar sightlines. 

The FLAMES spectra 
use the HR2 setting, providing a spectral resolution of $\sim$16 km\,s$^{-1}$ and wavelength coverage from $\sim$3850\AA\, to 4045\AA,
covering the Ca\,{\sc ii} K line. Full details of the sample, data reduction and analysis are given in Smoker et al. 
(2015) where the 
structure of the low-velocity (Galactic) gas and its variation on small scales is investigated and all the spectra 
are presented. The S/N ratios of the spectra are $\sim$30 or 60 per pixel for the two SMC clusters, and 95 or 135 for the two LMC clusters. 
Minimum and maximum star-star separations are 
11 arcsec to 27 arcmin, 14 arcsec to 20.7 arcmin, 14 arcsec to 22.2 arcmin, and 11 arcsec to 20 arcmin,
respectively, for NGC\, 330, NGC\, 346, NGC\,1761 and NGC\,2004. 
These correspond to transverse separations of $\sim$3--500\,pc at the distance of the Magellanic clouds. 

\begin{table*}
\caption{Basic data for the open clusters observed with FLAMES sorted in increasing NGC number. The distances to the 
LMC and SMC were taken from Keller \& Wood (2006). The ``Scales probed'' column corresponds to the minimum and 
maximum transverse star-star separation at the distance of the cluster. 
} 
\label{t_sample_FLAMES} 
\centering 
\begin{tabular}{lrrrrrrrrr} \hline
Cluster & Alternative & Location & ($l,b$) & Dist & Exp time & Median S/N & Stars & Scales  & Mag. scales \\
 & name & & (deg.) & (kpc) & (s) & at Ca\,{\sc ii} K   & used & probed ($^{'}$)  & probed (pc) \\
\hline 
NGC\,330 & Kron 35 & SMC & 302.42, --44.66 & 61 & 13650 & 30 & 110 & 0.2 -- 27.4 & 3.6 -- 486 \\
NGC\,346 & Kron 39 & SMC & 302.14, --44.94 & 61 & 6825 & 60 & 109 & 0.3 -- 20.7 & 5.3 -- 367 \\
NGC\,1761 & LH\,09 & LMC & 277.23, --36.07 & 51 & 13650 & 135 & 109 & 0.2 -- 22.3 & 3.0 -- 331 \\
NGC\,2004 & KMHK 991 & LMC & 277.45, --32.63 & 51 & 13650 & 95 & 75 & 0.2 -- 20.0 & 3.0 -- 297 \\
\hline 
\end{tabular}
\end{table*}

Additionally, FEROS\footnote{FEROS (Kaufer et al. 1999) is a
high-resolution echelle spectrograph, mounted on the 2.2\,m Telescope
at La Silla, Chile, operated by ESO.} and UVES 
observations towards 71 early-type stars in the LMC and 85 from the SMC were downloaded from the ESO archive. These observations 
have a spectral resolution of 6.3 km\,s$^{-1}$ with a median S/N ratio of 35 per pixel in Ca\,{\sc ii} and 75 in Na\,{\sc i} D. 
The stars observed are listed in Smoker et al. (2015), with 
Fig. \ref{f_MC_lb_FEROS} showing the location of the stars in ($l,b$) coordinates. 
The species considered in this paper are shown in Table \ref{t_wavef}.

\begin{figure}
\epsfig{file=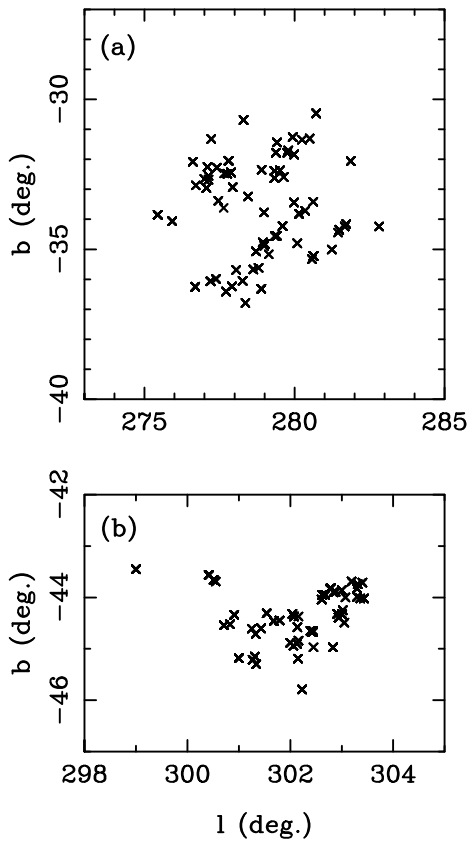}
\caption[]{Location of (a) LMC and (b) SMC stars observed with 
FEROS and UVES.} 
\label{f_MC_lb_FEROS}
\end{figure}

%
%
\begin{table}
\setcounter{table}{1}
\begin{center}
\caption[]{Main transitions studied in this paper. Wavelengths and
oscillator strengths are from Morton (2003, 2004). 
Column 4 gives the ionisation potential (IP) in eV. For comparison the
ionisation potential of H\,{\sc i} is 13.60 eV.}
\label{t_wavef}
\begin{tabular}{lrrr} 
\hline
Trans. & $\lambda_{\rm air}$ & $f$-value & IP \\
& (\AA) & & (eV) \\
\hline
Ca\,{\sc ii}  K  & 3933.661 & 0.627 & 11.87 \\
Na\,{\sc i}   D1 & 5889.951 & 0.641 & 5.14 \\
Na\,{\sc i}   D2 & 5895.924 & 0.320 & 5.14 \\
\hline
\end{tabular}
\end{center}
\end{table}

The data were reduced using the FEROS pipeline (in {\sc MIDAS}) or
the reduced spectra downloaded from the ESO archive in the case of UVES. For 
FEROS the reduction was undertaken 
using both standard and optimum extraction, with and without cosmic ray removal, respectively. 
Agreement was found to be good between the two methods. To check the quality of the results, 
the equivalent widths and velocities of a handful of lines were compared with previous 
UVES\footnote{UVES (Dekker et al. 2000) is a high
resolution echelle spectrograph, mounted on the VLT/Unit Telescope 2 (Kueyen)
at Cerro Paranal, Chile, operated by ESO.}
observations of a few B-type stars taken from the on-line version of the Paranal Observatory Project (POP) 
survey (Bagnulo et al. 2003). Agreement was found to be within 1 km\,s$^{-1}$ for velocities 
and within 5 per cent for equivalent widths for strong lines. 

The individual FEROS or UVES spectra were co-added 
using {\sc scombine} within {\sc iraf}\footnote{{\sc iraf}
is distributed by the National Optical Astronomy Observatories, U.S.A.}, converted into
{\sc ascii} format and then read into the spectral analysis software {\sc dipso} for further analysis. Initially this included shifting to the Kinematical 
Local Standard of Rest (LSR) using corrections generated by the program {\sc rv} (Wallace \& Clayton 1996), then 
normalising the spectra by fitting polynomials to the stellar continuum in the region of interest. 
The RMS of the normalisation procedure gives the S/N ratio in the final spectra. For the region around Na\,{\sc i} D, 
telluric spectra were removed as described in Hunter et al. (2006). 

Equivalent widths, velocity centroids and full width half maximum (FWHM) velocity values of the optical transitions were 
obtained by fitting Gaussians to the normalised spectra using the {\sc elf} package within {\sc dipso} (Howarth et al. 2003). 
These results were used as initial inputs to the {\sc vapid} code (Howarth et al. 2002) which provide the final 
values of the column densities and $b$-values by curve-of-growth analysis. Errors were derived as 
described in Hunter et al. (2006). 

\subsubsection{21\,cm data from the GASS and LAB surveys}
\label{s_LAB}

For the H\,{\sc i} 21\,cm spectra, we adopt measurements from the Parkes Galactic All-Sky Survey (GASS) and 
LABS survey (Kalberla et al. 2005; McClure-Griffiths et al. 2009). Both surveys have spectral 
resolutions of $\sim$1 km\,s$^{-1}$, with spatial resolutions of $\sim$0.5$^{\circ}$ and and 16 arcmin, respectively. 
The vast majority of these spectra show no I/HVC detection, hence an H\,{\sc i} map is not shown.
We note that H\,{\sc i} in I/HVCs is clumpy, with 
structures visible in H\,{\sc i} emission down to the observational limit of $\sim$1 arcmin in 
objects such as Complex C (Smoker et al. 2001), compact HVCs (de Heij, Braun \& Burton 2002), and 
miscellaneous IVCs (Ben Bekhti et al. 2009), and additionally in absorption down to scales of arcseconds or 
less for low-velocity gas (Diamond et al. 1989). Similarly, comparison of LMC hydrogen column 
densities derived from 21-cm H\,{\sc i} observations compared with Ly$\alpha$ show variations 
of a factor of 2--3 in some sightlines, indicating small-scale H\,{\sc i} structure (Welty et al. 2012). 
Hence there are large systematic uncertainties in the derived H\,{\sc i} to optical line ratios derived 
in the current work (see Wakker 2001 for a discussion).


\section{Results}
\label{Results}

\subsection{FLAMES-GIRAFFE Magellanic Cloud spectra in Ca\,{\sc ii}}
Figures \ref{f_MaxVariation_EW_I/HVCs_NGC_330_I/HVC_p45_p85} to \ref{f_MaxVariation_EW_I/HVCs_NGC_2004_I/HVC2_p75_p130} 
(available online) show, for each cluster, the spectra of the 16 star-to-star pairs with the maximum difference in equivalent 
width of the intermediate- or high- velocity component, to show the variation in Ca\,{\sc ii} K I/HVC absorption-line strength. 
Figure \ref{f_FLAMES_MinMax_EW_I/HVC} shows the corresponding 
plots for two or three objects with the strongest and weakest I/HVC component per cluster observed in Ca\,{\sc ii} K with FLAMES. 
All of the spectra are shown in Smoker et al. (2015), in which the low-velocity component only is discussed and 
where tables of the equivalent width measurements at all velocities are given. 

\begin{figure*}
\setcounter{figure}{2}
\includegraphics[]{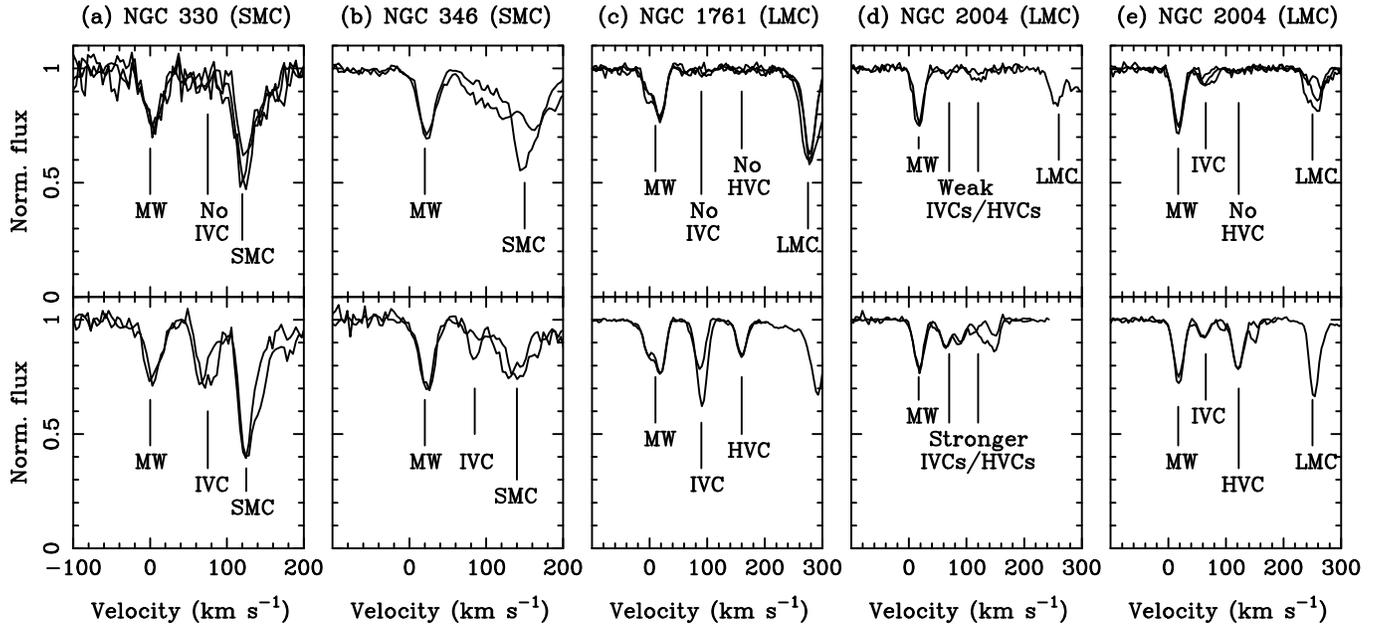}
\caption[]{Ca\,{\sc ii} K spectra towards the four Magellanic clusters studied, showing two or three sightlines with the 
minimum (top panels) and maximum (bottom panels) equivalent width in the I/HVC components. The maximum star-to-star separation 
on the sky is 27 arcminutes (the size of the FLAMES plate). It is
apparent that the variation in the strength of the low-velocity Ca\,{\sc ii} component
is smaller than the variation in the I/HVC and Magellanic components.} 
\label{f_FLAMES_MinMax_EW_I/HVC}
\end{figure*}

\subsection{FEROS and UVES Magellanic Cloud spectra in Ca\,{\sc ii}/Na\,{\sc i}} 

Smoker et al. (2015) present FEROS and UVES Ca\,{\sc ii} K and Na\,{\sc i} D1 spectra 
and Voigt profile fits of stars towards the Magellanic Clouds used in our analysis 
as well as the nearest GASS and LAB Survey H\,{\sc i} 21\,cm spectra. 


\section{Discussion}
\label{Discussion}

\subsection{Large-scale structure of I/HVCs toward the LMC}
\label{s_lss}

In this section we use the FEROS and UVES results to discuss the  velocity field towards the LMC IVCs, 
component structure observed in Ca\,{\sc ii}, variation with Ca\,{\sc ii} column density 
with position, and finally elemental abundances using the current optical observations and 
previous UV data taken from Lehner et al. (2009). 

\subsubsection{Velocity dependence on RA for LMC I/HVCs}

Due to its higher radial velocity, discriminating between HVC and Magellanic velocity components is easier for 
the LMC than for the SMC. Spectra for a total of 73 LMC stars exist in either Ca\,{\sc ii} or Na\,{\sc i} D, and HV components are present in 
many of them. In the LMC spectra there are I/HVC components present in Ca\,{\sc ii} K at a range of velocities from $\sim$+40 km\,s$^{-1}$ 
up to the LMC velocity of $\sim$+280 km\,s$^{-1}$. Lehner et al. (2009) find that the velocity of the HV components in the 
LMC standard of rest (LMCSR) correlates with Right Ascension, which they ascribe to the clouds being formed by an energetic 
outflow from the LMC. On the other hand, Richter et al. (2014) also find UV absorption-line profiles at high velocities 
some degrees away from the LMC, which is inconsitent with the outflow scenario and implies a separate origin.

\begin{figure}
\includegraphics[]{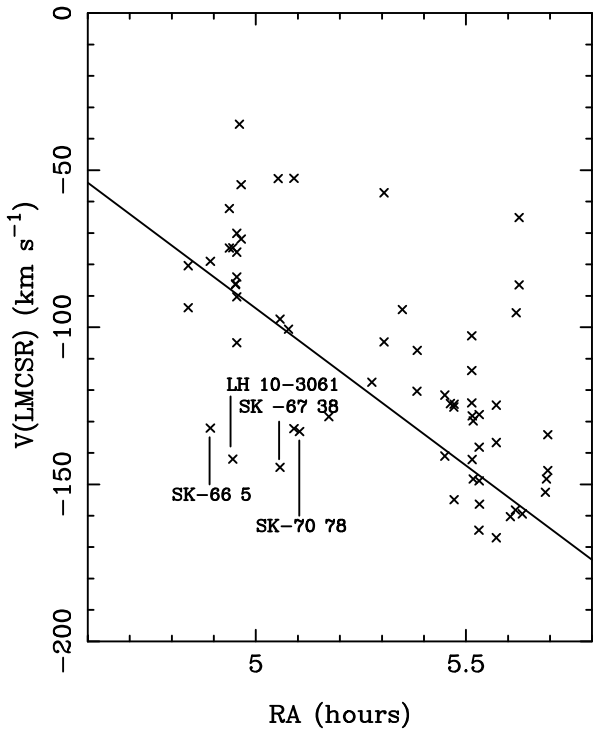}
\caption[]{Velocity of high velocity components in the LMC
standard of rest plotted against their RA. The solid line is the best-fit relationship of
Lehner et al. (2009). Only components with LSR velocities in the 
range +90--175 km\,s$^{-1}$ are shown.}
\label{f_LMCLSR_RA}
\end{figure}

\begin{figure}
\includegraphics[]{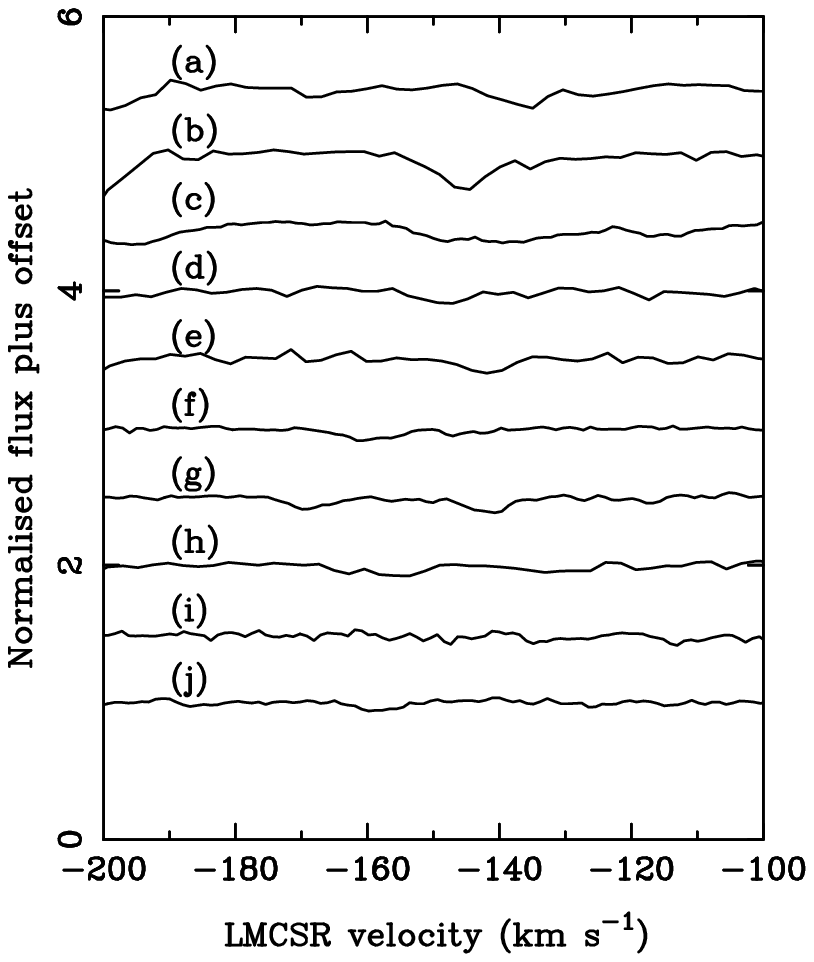}
\caption[]{Spectra for sightlines with components with LMC standard of rest less than 
--130 km\,s$^{-1}$. The ordinate is normalised flux plus an offset in incrments 
of 0.5. The abscissa is the LMCLSR. (a) SK-67 199 (b) SK-70 111 (c) HDE 269702 (d) SK-69 274 
(e) SK-66 118 (f) SK-67 38 (g) LH10 0361 (h) SK-66 5 (i) SK-70 69 (j) SK-66 171.}
\label{f_LMCLSR_Velminus130}
\end{figure}

The idea that some HVCs are in some way connected to the Magellanic system is not new (Giovanelli 1981; Mirabel 1981; 
Olano 2004 amongst others), although what fraction of them and the exact formation mechanism is still unclear (Nidever et al. 2008).
Following Lehner et al., Fig. \ref{f_LMCLSR_RA} shows the velocity, plotted against their RA,
of all detected components in Ca\,{\sc ii} K from --50 to --200 km\,s$^{-1}$ in the LMCSR, which is defined as:

\begin{equation}
v_{\rm LMCSR}=v_{\rm GSR} + (86 \times {\rm cos}l {\rm cos} b) + (268 \times {\rm sin}l {\rm cos}b) 
- (252 \times {\rm sin} b),
\label{LMCSR}
\end{equation}

where $v_{\rm GSR}$ = $v_{\rm LSR}$ + $v_{\odot}$sin$l$cos$b$ is the
velocity in the Galactic standard of rest frame, and $v_{\odot}$ is the solar 
circular rotation velocity around the Galactic centre. The solid line is the best-fit relationship 
of Lehner et al. (2009), and the current observations generally are offset 
by about +20 km\,s$^{-1}$ from this. The scatter in the two datasets is similar.
At RA=$\sim$05$^{\rm h}$30$^{\rm m}$ the range in $v$(LMCSR) is from $\sim$--100 to --170 km\,s$^{-1}$, 
rising to $\sim$--40 to --130 km\,s$^{-1}$ at RA$\sim$5$^{\rm h}$. A few stars (SK-66 5 and SK-70 78 
marked in Fig. \ref{f_LMCLSR_RA}) at RA$\sim$5$^{\rm h}$ show absorption features in their spectra 
that would be expected at higher values of RA. Figure \ref{f_LMCLSR_Velminus130} shows spectra 
that have LMC LSR velocities of less than --130 km\,s$^{-1}$. The features here are often close to 
the noise level.


\subsubsection{Velocity component structure in Ca\,{\sc ii} for LMC I/HVCs from FEROS observations}
I/HVC component structure is observed in some sightlines.  
For the IVCs these include SK-67 206, SK-67 173, SK-68 111, LHA 120-S 116 and
SK-70 120, while for the HVCs include SK-67 199 (although at low
S/N), SK-68 111, SK-69 43 and AzV 490 (although there is possible stellar
contamination). In only one case are there corresponding UV data from 
Lehner et al. (2009), specifically towards SK-68 111 where these authors only find a single component
with velocities of +109.6$\pm$3.8 km\,s$^{-1}$ (O\,{\sc i}), +102.4$\pm$1.4 km\,s$^{-1}$ (Fe\,{\sc ii}) and +114.5$\pm$3.8 km\,s$^{-1}$ (H\,{\sc i}), 
compared with our Ca\,{\sc ii} K velocities for HVC components of +96.1$\pm$1.1 and +114.6$\pm$1.2 km\,s$^{-1}$. 
The velocity resolution for the UV data is either 7 km\,s$^{-1}$ (STIS) or $\sim$20 km\,s$^{-1}$ (FUSE) and with typical S/N ratios 
of $\sim$5--40 in the FUSE spectra, with the velocity resolution in the Parkes data being 1.6 km\,s$^{-1}$ although
with a much larger beam than the optical observations. In any case, we note that no component structure is listed in Table 1 of Lehner et al. (2009) 
in either the UV or H\,{\sc i} data. However, an inspection of Fig. 1 of Lehner et al. (2009) indicates that at least for some 
sightlines such structure may be difficult to determine due to low S/N ratios. 

Multi-component velocity profiles have been detected in other I/HVCs
observed in absorption, 
including the Magellanic Bridge (Misawa et al. 2009), 
the Magellanic Stream (Fox et al. 2005, 2010), 
the M\,15 IVC (Meyer \& 
Lauroesch 1999;  Welsh, Wheatley \& Lallement 2009) and SN 1987A in the LMC (Adreani 
\& Vidal-Madjar 1988; Blades et al 1988a,b; Welty et al. 1999). 
This component structure indicates that several physical regions of absorbing gas exist 
along the line-of-sight, and suggests but does not prove the fragmentation of initially-larger clouds. 
Fragmentation is also observed in deep 21\,cm images of the tip of the
Magellanic Stream (Stanimirovi\'c et al. 2008), and is predicted by hydrodynamic 
simulations of HVCs streaming through a hot corona (e.g. Bland-Hawthorn 2009). Such 
fragmentation may be the precursor to eventual evaporation of the I/HVCs before they reach the 
disk of the Milky Way (e.g. Heitsch \& Putman 2009; Fox et al. 2010). A mix of fragmentation, 
cooling and mixing with other gas may change the ionisation structure of I/HVCs and 
their metalicities  (Gritton et al. 2014), with non-equilibrium chemistry in diffuse 
interstellar gas increasing cooling (Richings, Schaye \& Oppenheimer 2010) and perhaps 
explain the detection of Na\,{\sc i} in some I/HVCs.
If I/HVCs {\em do} evaporate before reaching the disk, then they must re-condense
if they are to form the fuel needed to sustain star formation in the disk and 
to reproduce stellar abundance patterns (Kennicutt 1998, Chiappini 2008). 

Finally, in eight of nine sightlines where we have measured the velocity of the {\em main} HVC component 
in Ca\,{\sc ii} and H\,{\sc i}, the values for both elements agree within the errors, providing 
evidence that at least in these cases the Ca\,{\sc ii} and H\,{\sc ii} sample the same phase of the 
interstellar medium. The discrepant case is towards SK-69 214 where the velocities are different 
by 5.4 km\,s$^{-1}$, although low S/N ratio in the H\,{\sc i} observations. 

\subsubsection{Variation of LMC I/HVC Ca\,{\sc ii} column density with position}

Figures \ref{f_logN_position_FEROS_1} and \ref{f_logN_position_FEROS_2}
show maps of the Ca\,{\sc ii} column density as a function of position for the I/HVCs observed 
towards the LMC, using the column densities derived from FEROS and UVES archival data. The log of the 
column density ranges from 
$<$10.7 cm$^{-2}$ (a 3$\sigma$ upper limit for SK-67\, 14 at $l,b$=278.27$^{\circ}$,--36.05$^{\circ}$) to 12.50 cm$^{-2}$ 
(SK-69\,214 at $l,b$=277.99$^{\circ}$,--31.84$^{\circ}$) for the integrated flux between +40 and +60 km\,s$^{-1}$, 
and from $<$10.7 cm$^{-2}$ (a 3$\sigma$ upper limit for SK-67 2 at $l,b$=278.36$^{\circ}$,--36.79$^{\circ}$) 
to 12.16 cm$^{-2}$ (SK-68\,114 at $l,b$=278.90$^{\circ}$,--32.35$^{\circ}$) for 
the integrated flux between +60 and +100 km\,s$^{-1}$. 

\begin{figure}
\includegraphics[]{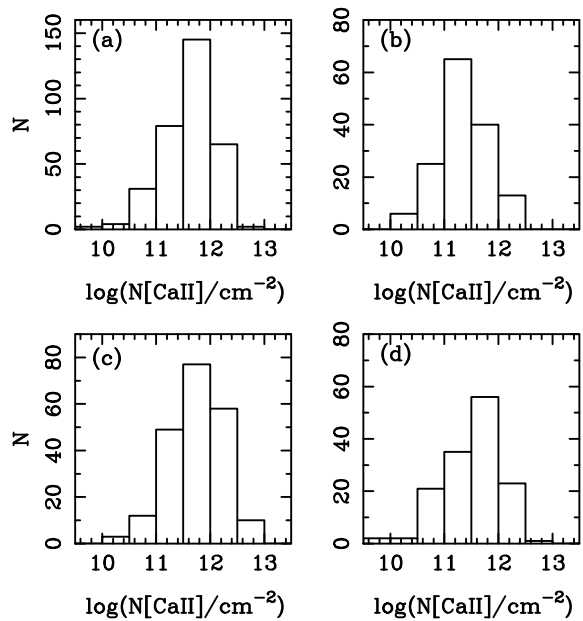}
\caption[]{Histogram of Ca\,{\sc ii} column densities for components derived from the FEROS and UVES spectra. 
Low-column density bins are severely affected by S/N ratio effects. 
(a) LV gas $v <$ +40 km\,s$^{-1}$ (b) +40$\le v \le$+100 km\,s$^{-1}$ (c) +100$\le v \le$+150 km\,s$^{-1}$ 
(d) $v >$ +150 km\,s$^{-1}$.}
\label{f_FEROS_CaK_N}
\end{figure}

Fig. \ref{f_FEROS_CaK_N} shows a histogram of the Ca\,{\sc ii} column densities observed towards the FEROS and UVES 
targets. For intermediate-velocity gas with LSR velocities between +40 and +100 km\,s$^{-1}$ the column density ranges 
from 10.94 to 12.50 dex with median value of 11.25 dex (N=50), compared with 11.0 to 12.4 (median=11.5 dex, N=18) 
in Ben Bekhti et al. (2008). For gas with velocities between +100 and +150 km\,s$^{-1}$ the range is from 
11.01 to 12.31 dex (median=11.51 dex), compared with 11.0 to 12.4 dex (median=11.5 dex, N=18) in Ben Bekhti et al. 
This compares with low-velocity Galactic gas in the current sample that has column densities ranging from 10.93 
to 12.58 dex with median value of 11.82 dex (N=97), compared with 10.9 to 12.6 (median=11.8 dex, N=16) for 
the corresponding velocities in Ben Bekhti et al., and median of 11.63 dex (N=362) towards low-velocity gas 
observed by Smoker et al. (2003).

\subsubsection{The Ca\,{\sc ii}/H\,{\sc i}, Ca\,{\sc ii}/Ca\,{\sc i}, Ca\,{\sc ii}/O\,{\sc i}, Ca\,{\sc ii}/Na\,{\sc i} 
and Na\,{\sc i}/H\,{\sc i} ratios for LMC I/HVCs} 

\begin{figure}
\includegraphics[]{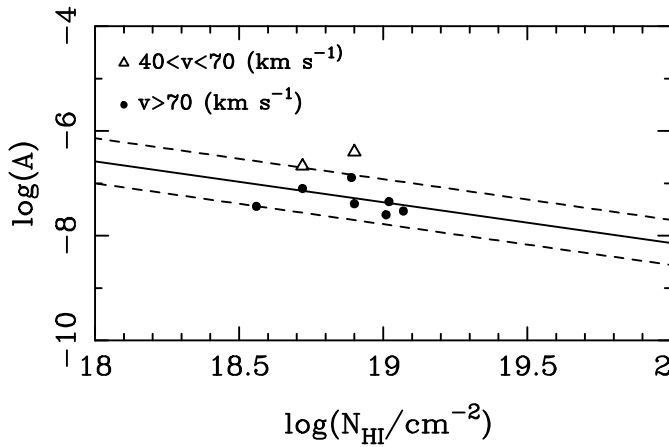}
\caption[]{Ca\,{\sc ii}/H\,{\sc i} ratio for FEROS and UVES sightlines with a detection in both species. The solid and dashed lines are the fit 
from Wakker \& Mathis (2000) and the associated RMS.}
\label{f_CaII_HI_Comparison}
\end{figure}

Table \ref{t_Lehner_Comparison} shows 21 stars for which more than 
one species was detected in Ca\,{\sc ii}, Na\,{\sc i}, O\,{\sc i}, Fe\,{\sc ii} or H\,{\sc i}. In 
only six of the sightlines are both Ca\,{\sc ii} and 21-cm H\,{\sc i} detected, with values 
of log\,[$N$(Ca\,{\sc ii})/$N$(H\,{\sc i})] ranging from --6.81 to --7.46. 
Five of the points lie on the best fit of I/HVCs studied by Wakker \& Mathis (2000). However, 
in the IVC toward SK-69 59, which has log\,$N$(H\,{\sc i})=18.89, the observed value of 
log\,[$N$(Ca\,{\sc ii})/$N$(H\,{\sc i})] of --6.89 is +0.39 dex higher than predicted by
Wakker \& Mathis, who found that generally halo gas has larger 
$N$(Ca\,{\sc ii})/$N$(H\,{\sc i}) ratios than disc gas. 
In any event, the nine data points shown in Fig. \ref{f_CaII_HI_Comparison}
follow the well-known trend that the gas-phase abundance of
Ca\,{\sc ii} decreases with increasing H\,{\sc i} column density, as
ions are removed from the gas phase and onto dust grains. In their sightline towards 
SN1987A, Welty et al. find slightly lower ratios of log(Ca\,{\sc ii}/H$_{\rm tot}$) = --7.2 to --7.8 for 
clouds with +100 to +225 km\,s$^{-1}$, with LMC features having corresponding ratios from --7.8 to --9.1.  
In our current sample we did not detect Ca\,{\sc i} in any of our FEROS sightlines, with S/N ratios 
of typically 50 to 90, with corresponding 5$\sigma$ column density upper limits of 
$\sim$10.5 to 10.2 dex. The ratio of Ca\,{\sc ii} to Ca\,{\sc i} therefore exceeds 1.5 
dex towards the HVC SK-69 59 and 2.1 towards IVC SK-69 214. 
Finally, we note that Lehner et al. (2009) find an average metallicity in their 139 LMC sightlines of 
[O\,{\sc i}/H\,{\sc i}]=--0.51, indicating a sub-solar metallicity for the HV gas and similar to 
that of the LMC which has Fe/H of $\sim$0.5 dex (e.g. Bertelli et al. 1992, Carrera et al 2008).

The Ca\,{\sc ii}/O\,{\sc i} ratio can be used to estimate the amount of dust present in the LMC HVCs. By studying two halo sightlines, 
Richter et al. (2009) found HVCs with $N$(Ca\,{\sc ii})/$N$(O\,{\sc i}) similar to the solar Ca/O abundance ratios, indicating that 
those clouds do not contain significant amounts of dust. However, in the diffuse ISM, depletion onto dust grains causes the 
value of Ca/O to be frequently more than 1.5 dex lower than the solar value. Comparing the current dataset to that of Lehner et al. (2009), we find 
values of N(Ca\,{\sc ii})/N(O\,{\sc i}) of $<$--3.39 for SK-67 256, 
--3.38 or --3.64 (depending on whether we count one or both Ca\,{\sc ii} components) for SK-68 111, 
--3.66 for HD\,269599, --3.06 for SK-69 50, --3.37 for SK-66 100, --3.46 for SK-67 168, --3.88 for SK-67 38, 
--3.09 for SK-70 60, --3.76 for SK-70 69 and 
--2.57 or --2.90 (depending on whether we count one or both Ca\,{\sc ii} components) for SK-70 78. 
Given the solar abundances (O/H)$_{\odot}$=--3.31 (Asplund et al. 2004) and (Ca/H)$_{\odot}$=--5.65 (Morton 2003, 2004),
the solar Ca/O ratio is --2.34, hence one of the current sightlines appears to be lightly depleted, with the other nine showing 
depletions of $\sim$1 dex. This suggests the presence of dust grains, although we have made no ionisation corrections, and 
in the diffuse ISM Ca\,{\sc ii} is a trace species with the majority of the calcium being in the form of Ca\,{\sc iii} 
(Sembach et al. 2000). Nonetheless, the presence of dust is also indicated in some of the sightlines studied by Lehner et al. (2009) 
that have sub-solar Si\,{\sc ii}/S\,{\sc ii} ratios, in agreement with the current result. 

\begin{figure*}
\includegraphics[]{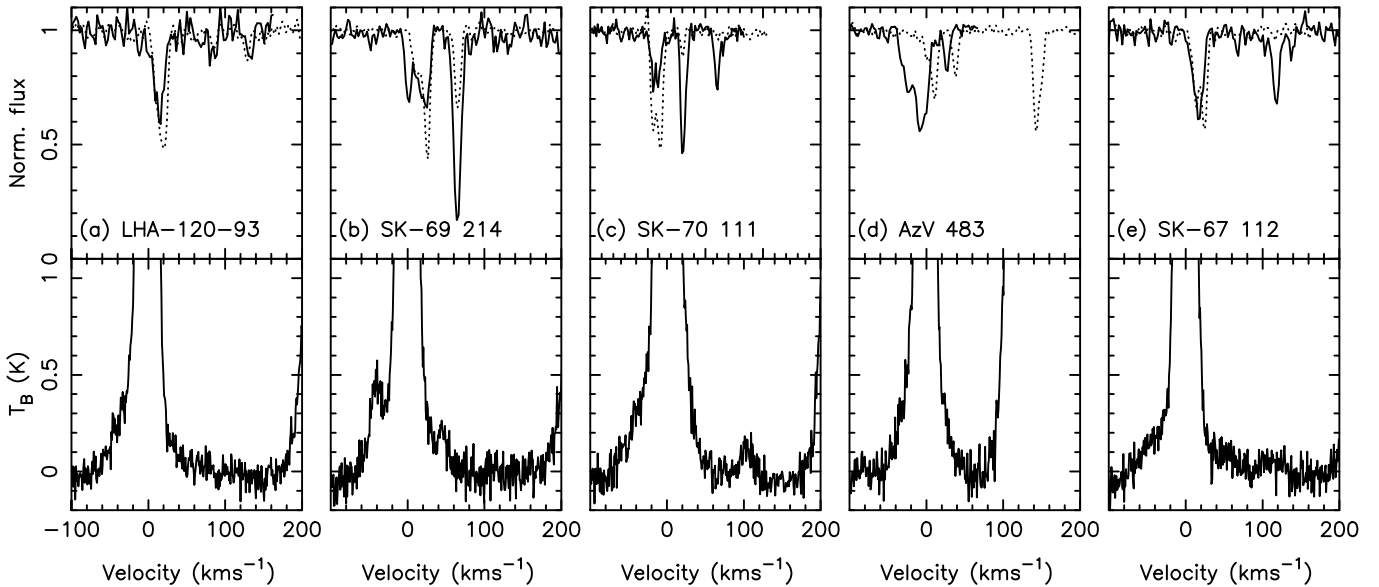}
\caption[]{(a)-(d) Ca\,{\sc ii} K, Na\,{\sc i} D and H\,{\sc i} LAB and GASS spectra 
of sightlines with possible Na\,{\sc i} I/HVC detections. (e) Corresponding spectra
with high Ca\,{\sc ii}/Na\,{\sc i} ratio. Filled lines in the top panel are 
Ca\,{\sc ii} with dashed lines being Na\,D1.}
\label{f_CaII_NaD_HI}
\end{figure*}

In common with earlier work (Routly \& Spitzer 1952, Siluk \& Silk 1974, Vallerga et al. 1993), 
the ratio of Ca\,{\sc ii}/Na\,{\sc i} D is found to increase markedly as one moves to high velocities, 
with the vast majority of sightlines only showing HVC absorption in Ca\,{\sc ii} K and not in Na\,{\sc i} D, 
although with some exceptions (e.g. Richter et al. 2009). In extragalactic sightlines the effect is 
less clear, with Richter et al. (2011) finding a range 
from --0.66 to 1.36, again typical of the diffuse warm interstellar medium. They note that in these 
conditions where the electron densities are less than $\sim$ 0.05 cm$^{-3}$, in dust-free gas the 
Ca\,{\sc ii}/Na\,{\sc i} ratio is roughly constant at +1.6 (c.f. Crawford 1992). In the current sightlines the 
velocities of the two species are similar, which may be taken as evidence that the two species originate within 
the same physical region. The same appears to be the case for the very few sightlines in the current sample 
that show both species in absorption, as displayed in Fig. \ref{f_CaII_NaD_HI}. In the current sample, the only 
detections of Na\,{\sc i} D at LSR velocities exceeding +40 km\,s$^{-1}$ (excluding the LMC and SMC) are: 
at $\sim$+68 km\,s$^{-1}$ with column density ratio log\,[$N$(Ca\,{\sc ii})/$N$(Na\,{\sc i})]= --0.44$\pm$0.05 dex) and $\sim$+110\,km\,s$^{-1}$ 
with column density ratio log\,[$N$(Ca\,{\sc ii})/$N$(Na\,{\sc i})]=+0.41$\pm$0.20 dex) towards LHA 120-S 93; a borderline detection at $\sim$+68\,km\,s$^{-1}$ 
towards SK-69 214 with log\,$N$(Na\,{\sc I})=10.38 cm$^{-2}$ but with no Ca\,{\sc ii} 
detected at the same velocity; at $\sim$+45\,km\,s$^{-1}$ towards SK-70 111 with log\,$N$(Ca\,{\sc ii})=12.11 cm$^{-2}$ 
and log\,$N$(Na\,{\sc i})=10.98 cm$^{-2}$ (a ratio of 1.13$\pm$0.07 dex); and finally at $\sim$+135 km\,s$^{-1}$ towards AzV 483 in the LMC. 
For high velocity gas, the maximum observed lower limit to the Ca\,{\sc ii}/Na\,{\sc i} ratio is towards SK-67 112, where 
log\,[$N$(Ca\,{\sc ii})/$N$(Na\,{\sc i})]$>$1.45 dex in the feature at 101.8 km\,s$^{-1}$. For intermediate velocity gas, 
the maximum ratio is toward SK-69 237 where log\,[$N$(Ca\,{\sc ii})/$N$(Na\,{\sc i})]$>$1.58 dex. 

For gas with absolute LSR velocities exceeding 40 km\,s$^{-1}$, Ben Bekhti et al. (2008) 
have five detections of both Ca\,{\sc ii} and Na\,{\sc i} D, which have Ca\,{\sc ii}/Na\,{\sc i} ratios 
of --0.2 (complex L), 0.4, 0.4 (Magellanic Stream), 0.5 and 1.4 dex (other gas). In extra-planar gas van Loon et al. (2009) 
find a ratio of +0.4 dex with a range from 0.23 to 0.56 dex, which is typical of low-velocity components 
measured by Welsh et al. (2009) and Smoker et al. (2015).  We note that for IVCs with a 
Ca\,{\sc ii}/Na\,{\sc i} ratio exceeding 5 the gas is characterised as having a temperature of $\sim$10,000 K 
and at least partially ionised (e.g Hobbs 1975, Welsh et al. 2009). Lower values of this ratio 
are found in denser and cooler clouds.


We only have upper limits to the Na\,{\sc i}/H\,{\sc i} ratios in the current sample. 
These are $\la$8.3 dex at an H\,{\sc i} column density of 19.0 dex, 
compared with the best-fit line from Wakker \& Mathis that predicts an 
abundance ratio of $\sim$8.1 dex at this H\,{\sc i} column density. This may
be related to beam-smearing for the H\,{\sc i} observations that we 
use. Higher S/N observations would likely detect Na\,{\sc i} in these sightlines.

We have decided not to include Cloudy because the Ca\,{\sc ii} and Na\,{\sc i} may 
not be physically associated with the H\,{\sc i}  (a fact that Cloudy assumes is the case), 
and hence the derived ionisation correction using a Cloudy model will likely be incorrect.
See Fox et al. (2013) for a discussion.


\begin{table*}
\small
\caption[]{Measurements of I/HVC absorption in sightlines in common with Lehner et al. (2009) from which the O\,{\sc i}, Fe\,{\sc ii} and H\,{\sc i} data are taken. Components
in Ca\,{\sc ii} K and Na\,{\sc i} D are shown when their velocity is greater than +40
km\,s$^{-1}$. Components from Lehner et al. (2009) are shown when
the velocity is from +90 to +175 km\,s$^{-1}$. Sightlines are ordered by increasing RA.} 
\label{t_Lehner_Comparison}
\begin{tabular}{lrrrrrrrrr}
\hline
Star        & $v$(O\,{\sc i})  & log$N$(O\,{\sc i})     & $v$(Fe\,{\sc ii}) & log$N$(Fe\,{\sc ii})  & $v$(H\,{\sc i})  & log$N$(H\,{\sc i}) & $v$(Ca\,{\sc ii}) & log$N$(Ca\,{\sc ii}) & log$N$(Na\,{\sc i}) \\  
            &   (km\,s$^{-1}$) &        (cm$^{-2}$)    &    (km\,s$^{-1}$) &        (cm$^{-2}$) &   (km\,s$^{-1}$) &    (cm$^{-2}$)  &  (km\,s$^{-1}$)   &   (cm$^{-2}$)     & (cm$^{-2}$ )     \\  
            &                  &                       &                   &                    &                  &                 &                   &                   &                  \\  
SK-69 43    &  --           & --                       & --                & --                 & 143.6$\pm$1.3    & 19.02$\pm$0.06  &142.7$\pm$0.3      & 11.67$\pm$0.11    &  $<$10.66        \\
SK-67 28    & --            & $<$14.28                 & --                & $<$13.19           & --               & $<$18.47        & --                & $<$11.32          & $<$10.26         \\  
SK-68 26    & --            & --                       & --                & $<$13.62           & --               & $<$18.47        & --                & $<$11.32          & $<$10.36         \\  
SK-69 59    & 175.9$\pm$0.6 & 15.06$\pm$0.02           & 173.4$\pm$1.3     & 13.79$\pm$0.03     & 172.4$\pm$2.5    & 18.89$\pm$0.11  & 170.4$\pm$0.4     & 12.00$\pm$0.02    & $<$10.56         \\  
SK-68 41    & --            & $<$13.92                 & --                & --                 & --               & $<$18.47        & --                & $<$11.56          & $<$10.48         \\  
SK-66 106   & --            & $<$14.63                 & --                & --                 & --               & $<$18.47        & --                & $<$11.38          & $<$10.36         \\  
SK-67 150   & --            & $<$14.39                 & --                & --                 & --               & $<$18.47        & --                & $<$11.16          & $<$10.38         \\  
SK-67 169   & --            & $<$14.11                 & --                & --                 & --               & $<$18.47        & 83.5$\pm$1.0      & 11.01             &  (stellar)       \\  
   "        & --            & --                       & --                & --                 & 117.4$\pm$1.8    & 18.93$\pm$0.08  &      --           & $<$10.89          & $<$10.24         \\
SK-70 78    &  97.2$\pm$4.4 & 14.27$\pm$0.15           & --                & --                 & --               & $<$18.44        & 78.5$\pm$4.9      & 11.43             & $<$10.38         \\  
   "        &  --           & --                       & --                & --                 & --               & "               & 91.4$\pm$0.5      & 11.37             & $<$10.38         \\  
HD\,269599  & 119.7$\pm$3.4 & 15.20$\pm$0.08           & 110.6$\pm$1.6     & 13.99$\pm$0.05     & 119.3$\pm$5.2    & 19.07$\pm$0.09  & 117.4$\pm$0.7     & 11.54$\pm$0.03    & $<$10.56         \\  
SK-71 42    &  --           & --                       & --                & --                 &  66.8$\pm$2.3    & 18.72$\pm$0.14  & 66.0$\pm$0.2      & 12.05$\pm$0.04    & $<$10.45         \\
LHA-120 116 &  --           & --                       & --                & --                 & 114.9$\pm$1.2    & 19.11$\pm$0.07  &115.9$\pm$1.1      & 11.66$\pm$0.18    & $<$10.49         \\
SK-68 111   & 109.6$\pm$3.8 & 15.15$\pm$0.11           & 102.4$\pm$1.4     & 14.35$\pm$0.05     & --               & --              & 96.1$\pm$1.1      & 11.41$\pm$0.03    & $<$10.48         \\  
   "        &  --           & --                       & --                & --                 & 114.54$\pm$3.8   & 19.01$\pm$0.06  & 114.6$\pm$1.2     & 11.51$\pm$0.03    & $<$10.48         \\  
SK-69 214   &  --           & --                       & --                & --                 &  43.9$\pm$2.0    & 18.90$\pm$0.09  & 49.3$\pm$0.2      & 12.50$\pm$0.06    & 11.60$\pm$0.02   \\
SK-69 274   &  --           & --                       & --                & --                 & 100.1$\pm$3.2    & 18.56$\pm$0.20  &102.3$\pm$0.8      & 11.12$\pm$0.08    & $<$10.36         \\
SK-70 111   &  --           & --                       & --                & --                 & 103.1$\pm$1.9    & 18.72$\pm$0.11  &104.6$\pm$0.4      & 11.62$\pm$0.06    & $<$10.36         \\
            &  --           & --                       & --                & --                 & --               &  --             &116.5$\pm$1.9      & 11.10$\pm$0.29    & $<$10.36         \\   
SK-67 256   & 144.7$\pm$1.7 & 14.95$\pm$0.03           & 138.6$\pm$1.4     & 13.67$\pm$0.04     & 141.0$\pm$6.0    & 18.64$\pm$0.13  & --                & $<$11.56          &  (stellar)       \\  
SK-68 171   & --            & $<$14.46                 & --                & --                 & --               & $<$18.47        & 79.3$\pm$0.4      & 11.66             & $<$10.56         \\  
SK-70 120   & --            & --                       & --                & $<$13.36           & --               & $<$18.47        & --                & $<$11.32          & $<$10.56         \\  
 SK-68 63   & --            & --                       & --                & --                 & --               & --              &  100.4$\pm$0.1    &   11.17$\pm$0.17  & $<$10.19         \\
      "     & --            & --                       & --                & --                 & --               & --              &  136.1$\pm$0.7    &   10.38$\pm$0.18  & $<$10.19         \\
      "     & --            & --                       & --                & --                 & --               & --              &  152.1$\pm$0.2    &   11.10$\pm$0.17  & $<$10.19         \\
SK-65 47    & --            & --                       & --                & $<$12.80           & --               & $<$18.47        &   47.9$\pm$0.6    &   11.39$\pm$0.17  & $<$10.36         \\
SK-65 47    & 144.1$\pm$4.9 & 14.04$\pm$0.10           & --                & $<$12.80           & --               & $<$18.47        &  145.7$\pm$0.4    &   10.94$\pm$0.17  & $<$10.36         \\
BI 128      & --            & $<$14.49                 & 130.8$\pm$2.5     & 14.04$\pm$0.05     & --               & $<$18.61        &  132.1$\pm$0.8    &  11.06$\pm$0.30   & $<$10.66         \\
   "        & --            & $<$14.49                 & --                & --                 & --               &                 &  178.7$\pm$1.7    &  11.06$\pm$0.30   & $<$10.66         \\
SK-66 18    & --            & --                       & --                & $<$13.29           & --               & $<$18.47        &  165.6$\pm$0.0    &  11.51$\pm$0.17   & $<$10.34         \\
SK-66 100   & 107.3$\pm$3.1 & 14.59$\pm$0.06           & 109.6$\pm$1.4     & 13.74$\pm$0.04     & --               & $<$18.48        &  109.2$\pm$1.7    &  11.22$\pm$0.18   & $<$10.34         \\
            & --            & --                       & --                & --                 &                  &   --            &  120.4$\pm$0.4    &  10.87$\pm$0.18   & $<$10.34         \\
SK-67 168   & 110.5$\pm$2.6 & 15.00$\pm$0.06           & 114.5$\pm$1.3     & 14.13$\pm$0.03     & --               & $<$18.55        &  108.9$\pm$0.5    &  11.54$\pm$0.17   & 10.87$\pm$0.18   \\
      "     & --            & --                       & --                & --                 & --               &   --            &  119.3$\pm$0.6    &  10.74$\pm$0.17   & $<$10.49         \\
            & --            & --                       & --                & --                 & --               &   --            &  132.3$\pm$0.3    &  11.36$\pm$0.18   & $<$10.49         \\
BI 237      & --            & --                       & 109.6$\pm$3.0     & 14.01$\pm$0.05     & --               & $<$18.62        &  103.6$\pm$0.8    &  11.29$\pm$0.17   & $<$10.58         \\
            & --            & --                       & --                & --                 & --               &    --           &  131.1$\pm$0.6    &  11.21$\pm$0.17   & $<$10.58         \\
SK-67 38    & 125.2$\pm$2.9 & 14.23$\pm$0.10           & 128.1$\pm$2.0     & 13.47$\pm$0.08     & --               & $<$18.44        &  126.4$\pm$1.0    &  10.35$\pm$0.30   & $<$10.58         \\
SK-67 05    & --            & --                       & 112.3$\pm$3.6     & 13.79$\pm$0.06     & --               & $<$18.63        &  119.2$\pm$0.9    &  10.25$\pm$0.17   & --               \\
      "     & --            & --                       & --                & --                 & --               &   --            &  132.5$\pm$1.0    &  10.89$\pm$0.17   & --               \\
BI 253      & --            & --                       & --                & $<$13.36           & --               & $<$18.47        &  107.1$\pm$1.0    &  11.05$\pm$1.07   & $<$10.49         \\
SK-68 52    & --            & $<$14.26                 & --                & $<$13.12           & --               & $<$18.47        &  --               &  --               & --               \\
SK-69 50    & 147.9$\pm$0.9 & 15.42$\pm$0.02           & 145.3$\pm$1.2     & 14.39$\pm$0.03     & 146.9$\pm$3.8    & 19.23$\pm$0.10  &  142.1$\pm$0.9    &  11.62$\pm$0.18   & $<$10.55         \\ 
            & --            & --                       & --                & --                 &  --              &   --            &  148.1$\pm$0.0    &  11.38$\pm$0.17   & $<$10.55         \\
SK-70 60    & 123.3$\pm$1.1 & 14.64$\pm$0.07           & 128.7$\pm$1.7     & 13.79$\pm$0.06     & --               & $<$18.45        &  122.7$\pm$0.3    &  11.55$\pm$0.17   & --               \\     
SK-70 69    & 120.5$\pm$2.9 & 14.71$\pm$0.07           & 108.7$\pm$1.7     & 13.63$\pm$0.07     & --               & $<$18.48        &  111.9$\pm$0.6    &  10.95$\pm$0.18   & $<$10.49         \\
\hline 
\end{tabular}
\normalsize
\end{table*}

\subsubsection{FEROS and UVES SMC sightlines}
Eighty seven sightlines are present in the current sample towards the
SMC. As the SMC has a recessional LSR velocity of $\sim$+160 km\,s$^{-1}$,
the separation of the HVC and SMC components is not as marked as in the LMC. 
The Ca\,{\sc ii} column density ranges from $<$10.7 cm$^{-2}$ to 11.6 cm$^{-2}$ for the integrated flux between 
+60 and +100 km\,s$^{-1}$.

\subsection{Small-scale spatial structure of I/HVCs towards the LMC and SMC from FLAMES-GIRAFFE observations}
\label{s_sss}

\begin{figure*}
\includegraphics[]{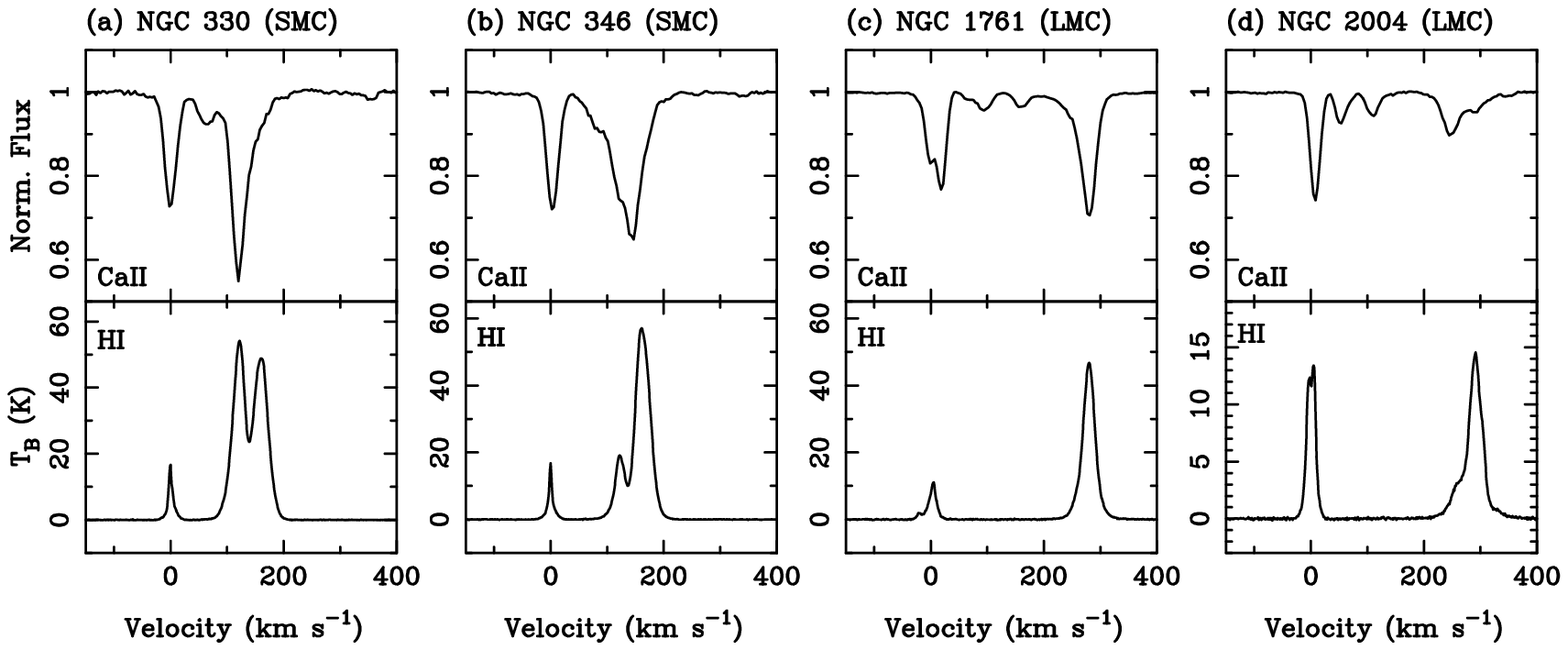}
\caption[]{Ca\,{\sc ii} K composite GIRAFFE spectra and H\,{\sc i}
GASS and LABS survey spectra for four clusters. See Sect. \ref{s_sss} for details.}
\label{f_LMC_SMC_cluster_composite}
\end{figure*}

Non-Magellanic Ca\,{\sc ii} K absorption is detected at intermediate- or high-velocities in many sightlines. 
The minimum star-star distance is 11 arcsec and the maximum is 27 arcminutes. Figure \ref{f_LMC_SMC_cluster_composite} shows the 
{\em composite} Ca\,{\sc ii} K spectra towards the four clusters, formed by median-combining the individual normalised spectra. 
The spectra have been boxcar smoothed using a box of three pixels, which  
retains the full spectral resolution as the FWHM of emission features measured from the arc frames is four pixels. 
For NGC\,330 the composite spectrum S/N ratio in the boxcar smoothed spectrum is $\sim$500, being $\sim$1100 for NGC\,346, 
$\sim$1200 for NGC\,1761 and $\sim$1200 for NGC\,2004. Also shown in Fig. \ref{f_LMC_SMC_cluster_composite} are the 
GASS and LABS survey 21\,cm H\,{\sc i} spectrum towards both clusters. These data are presented simply to confirm the weakness in H\,{\sc i} 
of the I/HVC emission -- in the GASS and LAB spectra no such emission is seen. 

\begin{figure}
\includegraphics[]{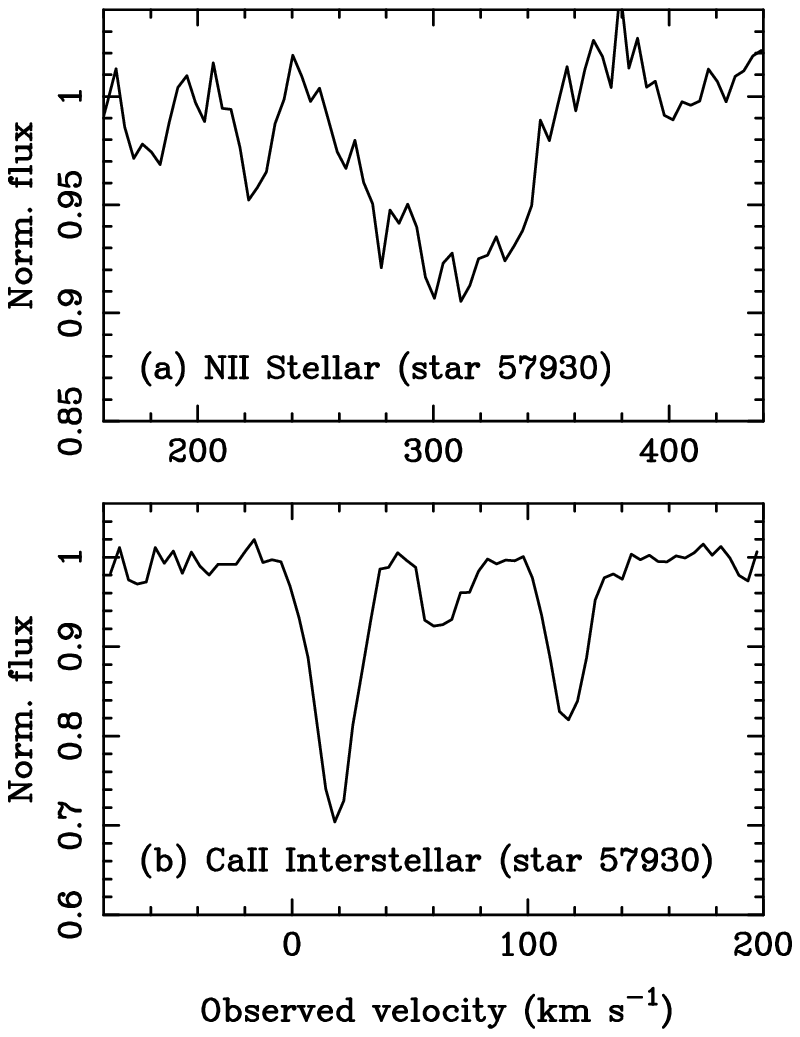}
\caption[]{FLAMES-GIRAFFE spectra of star 57930 in NGC\,2004 (LMC). (a) Stellar absorption line
of N\,{\sc ii} (3995\AA) with FWHM$\sim$60 km\,s$^{-1}$ (b) Absorption
lines around Ca\,{\sc ii} K (3933\AA). The profiles uncorrected for
instrumental broadening of $\sim$16 km\,s$^{-1}$ have
FWHM$\sim$17--19 km\,s$^{-1}$, implying that the lines are
narrow (possibly unresolved) and hence interstellar or circumstellar rather than stellar.} 
\label{f_I/HVC_is_or_stellar}
\end{figure}

An immediate question to ask is whether the observed absorption at I/HVC velocities could be stellar in nature. 
Figure \ref{f_I/HVC_is_or_stellar} 
shows an example of a star in NGC\,2004 where strong I/HVC Ca\,{\sc ii}
absorption is detected at observed (raw) velocities of +63 and +116
km\,s$^{-1}$. The FWHM of these lines is $\sim$17 km\,s$^{-1}$ (most or all of
which is caused by instrumental broadening of $\sim$16 km\,s$^{-1}$),
compared to the width of the N\,{\sc ii} stellar line at 3995\AA\ of
$\sim$60 km\,s$^{-1}$. The narrowness of these I/HVC components 
implies they are circumstellar or interstellar rather than stellar.
Furthermore, the fact that I/HVC features are observed at similar
LSR velocities in both the Ca\,K and H\,{\sc i} lines where the latter is detected 
(Figs. \ref{f_NGC_1761_GIRAFFE_TwoComponentVelCaK}  and \ref{f_NGC_2004_GIRAFFE_TwoComponentVelCaK})  
toward many different stars is very difficult to explain if
these lines are circumstellar, since in that case the velocities would
vary with the velocity of the star.

We now briefly discuss the Ca\,{\sc ii} K absorption line components 
toward the four individual Magellanic Cloud clusters observed with FLAMES-GIRAFFE. 

\subsubsection{NGC\,330 (SMC)}
The sightlines toward NGC\,330 (Figs. \ref{f_FLAMES_MinMax_EW_I/HVC}a and \ref{f_FLAMES_NGC_330_IVC_EW_Vel}) in the SMC show 
Ca\,{\sc ii} absorption at intermediate velocities between $\sim$+60--80 km\,s$^{-1}$, well 
separated from both the low velocity and SMC gas. Peak Ca\,{\sc ii} equivalent widths at IVC velocities are $\sim$80\,m\AA, 
with many sightlines having values of $\sim$60\,m\AA. A number of sightlines show {\em no} corresponding 
Ca\,{\sc ii} absorption with a S/N ratio of $\sim$60, which following Eqn. \ref{eqlim} gives a 3$\sigma$ upper limit on the equivalent width 
of $\sim$10\,m\AA, given that the instrumental resolution ($\Delta\lambda_{\rm instr}$) is 0.21\AA \, for the GIRAFFE HR2 setting 
and assuming an unresolved component, viz:

\begin{equation}
EW_{\rm lim}({X})=3\times(\rm S/N)_{\rm cont}^{-1} \Delta\lambda_{\rm instr}, 
\label{eqlim}
\end{equation}

Hence the variation in the Ca\,{\sc ii} equivalent width for IVC
components exceeds a factor of $\sim$8 ($\sim$0.9 dex) over the face of the cluster.
The velocity map in Fig. \ref{f_FLAMES_NGC_330_IVC_EW_Vel} shows no obvious large-scale structure or 
gradients, although clumps of gas at higher central velocities are for example present 
at ($l,b$)$\sim$(302.5$^{\circ}$,--44.6$^{\circ}$) with central velocities some $\sim$20 km\,s$^{-1}$
higher than towards the cluster core at ($l,b$)$\sim$(302.4$^{\circ}$,--44.7$^{\circ}$) which 
is a transverse distance of $\sim$150 pc at the distance of the SMC. 

\subsubsection{NGC\,346 (SMC)}
Although I/HVC absorption is visible in many of the FLAMES NGC\,346 sightlines 
(Fig. \ref{f_FLAMES_MinMax_EW_I/HVC}b), the gas merges into the 
SMC material. Hence we do not comment on the variation in these components. The highest velocity gas (presumably tracing 
the SMC itself) shows a huge variation 
in absorption-line strength and number of components. This is likely  explained by 
the cluster having a finite depth within the SMC. 

\subsubsection{NGC\,1761 (LMC)}
\begin{figure}
\includegraphics[]{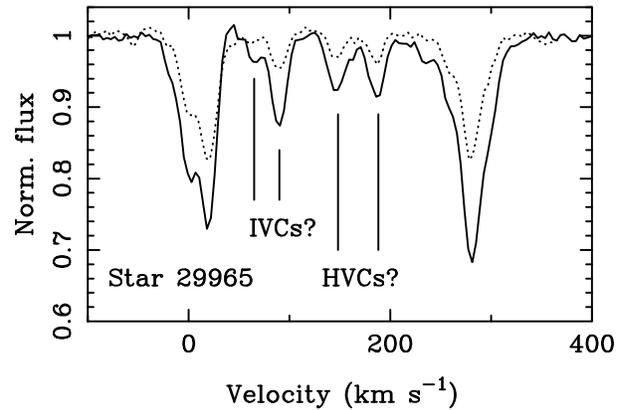}
\caption[]{Two-component FLAMES-GIRAFFE Ca\,{\sc ii} absorption-line structure in the I/HVCs towards NGC\,1761 in the LMC. The full line is for 
Ca\,K and the dotted line for Ca\,H.}
\label{f_NGC_1761_GIRAFFE_TwoComponentVelCaK}
\end{figure}

NGC\,1761 (LH09) is an LMC cluster that displays two apparently discrete Ca\,{\sc ii} absorption-line 
components in its interstellar spectra, well-separated in velocity, which are not at Milky Way or LMC velocities 
(Figs. \ref{f_FLAMES_MinMax_EW_I/HVC}c, \ref{f_FLAMES_LH09_v_p30_p105_I_HVCs_EW_Vel} and 
\ref{f_Fin_Prof_Coldns_Results_FLAMES_LH09_LMC1_p125_p200}). One is an IVC at $\sim$+90 km\,s$^{-1}$ with a maximum equivalent width of $\sim$100\,m\AA, and the other 
is an HVC at $\sim$+160 km\,s$^{-1}$ with a maximum equivalent width of $\sim$40\,m\AA. One sightline 
shows no Ca\,{\sc ii} absorption at either of these two velocities, in
data with a S/N ratio of $\sim$90, corresponding to a 3$\sigma$ 
equivalent width upper limit of 7\,m\AA.
Thus the Ca\,{\sc ii} equivalent widths vary across the cluster by factors of
$>$14 (IVC gas) and $>$6 (HVC gas).
NGC\,1761 is hence the cluster in which the biggest variation in Ca\,{\sc ii} absorption-line strength is seen 
in the current sample. As in the FEROS and UVES data, there are hints of two-component velocity structure towards a 
handful of sightlines, examples being shown in Fig. \ref{f_NGC_1761_GIRAFFE_TwoComponentVelCaK}. 
The velocities of the gas in Fig. \ref{f_FLAMES_LH09_v_p30_p105_I/HVCs_EW_Vel} for the IVC 
between +30 and +105 km\,s$^{-1}$ show a transverse gradient of $\sim$20 km\,s$^{-1}$ moving from east to 
west in Galactic longitude. Between +125 and +200 km\,s$^{-1}$, two clumps of gas are present 
(c.f. Fig. \ref{f_NGC_1761_GIRAFFE_TwoComponentVelCaK}, with the lower velocity clump to the East 
centred on $l,b$$\sim$277.1$^{\circ}$,--36.05$^{\circ}$) having velocities from $\sim$+140--150 km\,s$^{-1}$ and 
the structure to the west showing a larger variation in velocity. 

\subsubsection{NGC\,2004 (LMC)}
\begin{figure}
\includegraphics[]{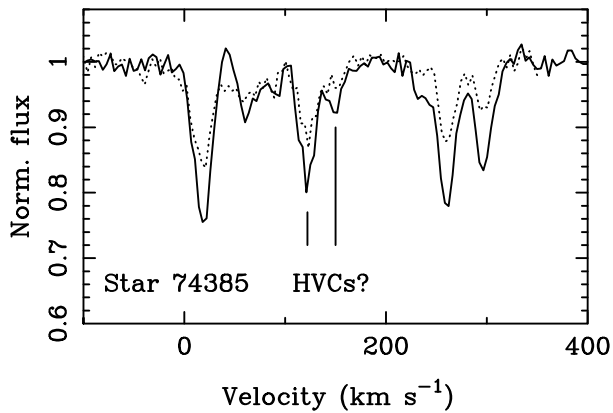}
\caption[]{Two-component FLAMES-GIRAFFE Ca\,{\sc ii} absorption-line structure in the I/HVCs towards NGC\,2004 in the LMC. The full line is for 
Ca\,K and the dotted line for Ca\,H.}
\label{f_NGC_2004_GIRAFFE_TwoComponentVelCaK}
\end{figure}

NGC\,2004 in the LMC shows the most complexity in intermediate- and high-velocity gas among the four clusters studied, with 
at least three I/HVC components visible in Ca\,{\sc ii}, at $\sim$+60, $\sim$+120 and $\sim$+150 km\,s$^{-1}$, the latter two components 
sometimes being merged at the FLAMES resolution (Figs. \ref{f_FLAMES_MinMax_EW_I/HVC}d, \ref{f_FLAMES_MinMax_EW_I/HVC}e, 
\ref{f_FLAMES_NGC2004_v_p30_p75_EW_Vel} and \ref{f_Fin_Prof_Coldns_Results_FLAMES_NGC2004_LMC1_p75_p130}). The peak equivalent widths 
in the first two components are 36 and 63\,m\AA, which are $\sim$5 and $\sim$9 times stronger than the 3$\sigma$ upper 
limits in the cases where no I/HVC gas is detected in Ca\,{\sc ii} absorption. As in the FEROS and UVES data, there are hints 
of two-component velocity structure towards a few sightlines, an examples being shown 
in Fig. \ref{f_NGC_2004_GIRAFFE_TwoComponentVelCaK}. 


\subsubsection{Summary of observed I/HVC EW variations}

EW variations in Ca\,{\sc ii} for  Magellanic Cloud I/HVCs exceed a factor of 10 (or $\sim$1 dex in column density) 
on transverse scales as small as $\sim$5 pc, assuming that the clouds lie at a distance of $\sim$55 kpc. If the 
clouds are in the halo of the Milky Way then the transverse scales would be reduced accordingly. 
These variations are large but not compared to previous work on the M\,15 IVC by Meyer \& Lauroesch (1999)  
and Welsh et al. (2009), who found column density variations of a factor exceeding 10 in IVC gas on scales of $<$0.1 pc. In 
the Galactic ISM, variations of up to 2 dex are also present on AU to pc scales in Na\,{\sc i} (Welty \& Fitzpatrick 2001; 
Points, Lauroesch \& Meyer 2004; Lauroesch 2007 and references therein; van Loon et al. 2009; Welsh et al. 2009  amongst others). 
Similarly, Appendix 15 of Wakker (2001) assumes a factor 2 in variation caused by H\,{\sc i} small-scale structure, 
factors of 1.5 (Ca\,{\sc ii}) and 2.5 (Na\, {\sc i}) to account for variations in depletion and further factors of 2 (Ca\,{\sc ii}) 
and 6 (Na\,{\sc i}) for ionisation variations. Typically, then, pc-scale variations in Ca\,{\sc ii} and Na\,{\sc i} are of the order 
0.8 dex and 1.5 dex, respectively, which is in line with the current work. The fact that in Milky Way extra planar gas, these variations 
exist in material far away from supernova remnants, has been taken to imply that this small-scale structure is either continuously 
regenerated or persists for long periods of time (e.g. van Loon et al. 2009, although see Marasco \& Fraternali 2011).

We have analysed the variations in Ca\,{\sc ii} equivalent width towards three clusters observed 
with FLAMES/GIRAFFE in a total of five velocity ranges.
Figure \ref{EWvariation} shows the percentage difference in equivalent width 
plotted against the distance between the stellar sighlines, with the derived upper limits shown 
in open circles with the measurements being filled circles. Table \ref{t_smallscalevariation} shows the
derived mean, median and standard deviation of the variaion in the equivalent width values at a range
of velocities. Towards NGC\,2004 the variation in the high velocity component is typically a 50 to 
100 percent larger than to the intermediate velocity component. This is naturally explained by the 
HVC being further away than the IVC and hence only the larger scales being sampled in the HVC. 
However, the same behaviour is not seen towards the I/HVC components towards NGC\,1761. Likewise, 
there is no clear increase in scatter in the equivalent width for any of the clusters as the 
scales sampled increase from 0.05 to 0.4 degrees. 

FLAMES-GIRAFFE observations of the Tarantula nebula have been presented in Na\,{\sc i} by van Loon et al. 
2013, who find that the standard deviation in equivalent width in the LMC gas is of the order of 25 percent, 
being 7 per cent for the Milky Way gas. The relative variation in Na\,{\sc i} absorption as a function of
angular sky separation increases from $\sim$0.50 at 10 arcsecond separation to 0.65 at 100 arcseconds, although 
with correlation coefficient of only 0.27.

Similarly, Van Loon et al. (2009) and Smoker et al (2015) found variations in the Ca\,{\sc ii} equivalent 
width in low velocity gas of $\sim$10 per cent on scales of $\sim$0.1, a factor of $\sim$10 smaller 
than seen in the current I/HVC sightlines. These fluctuations were explained by a simple model 
of the ISM comprised of spherical clouds of radii between 1 AU and 10-pc with filling factor 
of $\sim$0.2 (van Loon et al. 2009 Appendix B). Due to the limited spatial resolution of our study (scales 
probed from around 5-pc to 500-pc assuming the clouds are at the distance of the Magellanic 
Clouds) we cannot use these observations to say much about the size of the clouds although 
from the lack of detection of I/HVC in adjacent FLAMES-GIRAFFE fibres it is clear that the 
filling factor is somewhat less than unity. 

\begin{table*}
\setcounter{table}{3}
\begin{center}
\caption[]{Percentage variation in observed Ca\,{\sc ii} equivalent width for I/HVCs in the FLAMES sample 
as a function of star-to-star separation for three Magellanic clusters.}
\label{t_smallscalevariation}
\begin{tabular}{lrrrrr} 
\hline
Cluster     & velocity range &   Separation  & EW$_{\rm Med}$ & EW$_{\rm Ave}$ & EW$_{\rm sigma}$  \\
            &(km\,s$^{-1}$)   &   (arcmin)    &  variation    & variation    &   variation       \\
            &                &               &  (percent)     &  (percent)  &  (percent)     \\
\hline
NGC\,330    &    45 to 85    &  0.05 to 0.10  &    64.2       &   98.0        &     104.4         \\
NGC\,1761   &    60 to 120   &        "       &    63.2       &  101.4        &     118.5         \\
NGC\,1761   &   130 to 190   &        "       &    69.6       &  127.6        &     156.5         \\
NGC\,2004   &    40 to 100   &        "       &    51.6       &   78.0        &      83.0         \\
NGC\,2004   &   100 to 180   &        "       &   108.4       &  174.8        &     193.9         \\
NGC\,330    &    45 to 85    &  0.10 to 0.15  &    63.4       &  103.9        &     121.4         \\
NGC\,1761   &    60 to 120   &        "       &    73.1       &  121.8        &     158.2         \\
NGC\,1761   &   130 to 190   &        "       &    97.4       &  148.7        &     161.7         \\ 
NGC\,2004   &    40 to 100   &        "       &    60.5       &   90.6        &      94.4         \\
NGC\,2004   &   100 to 180   &        "       &    56.7       &  108.4        &     193.0         \\
NGC\,330    &    45 to 85    &  0.15 to 0.20  &    71.7       &  113.5        &     133.6         \\
NGC\,1761   &    60 to 120   &        "       &    76.8       &  139.4        &     189.2         \\
NGC\,1761   &   130 to 190   &        "       &    98.6       &  146.9        &     165.1         \\
NGC\,2004   &    40 to 100   &        "       &    64.7       &  100.1        &     136.7         \\
NGC\,2004   &   100 to 180   &        "       &   110.2       &  179.7        &     216.0         \\
NGC\,330    &    45 to 85    &  0.20 to 0.25  &    75.4       &  102.4        &     100.7         \\
NGC\,1761   &    60 to 120   &        "       &   112.0       &  172.6        &     199.1         \\
NGC\,1761   &   130 to 190   &        "       &    88.7       &  144.3        &     148.7         \\
NGC\,2004   &    40 to 100   &        "       &    64.2       &  111.2        &     136.7         \\
NGC\,2004   &   100 to 180   &        "       &   112.3       &  183.2        &     236.7         \\
NGC\,330    &    45 to 85    &  0.25 to 0.30  &    49.1       &  102.2        &     190.4         \\
NGC\,1761   &    60 to 120   &        "       &   124.7       &  211.2        &     240.6         \\
NGC\,1761   &   130 to 190   &        "       &    95.3       &  156.6        &     182.1         \\
NGC\,2004   &    40 to 100   &        "       &   109.8       &  151.1        &     159.0         \\
NGC\,2004   &   100 to 180   &        "       &   104.7       &  157.5        &     176.5         \\
NGC\,330    &    45 to 85    &  0.30 to 0.35  &    42.2       &   72.1        &      78.1         \\
NGC\,1761   &    60 to 120   &        "       &   236.2       &  348.2        &     148.7         \\
NGC\,1761   &   130 to 190   &        "       &    92.0       &  123.5        &      98.3         \\
NGC\,2004   &    40 to 100   &        "       &   163.5       &  224.0        &     182.0         \\
NGC\,2004   &   100 to 180   &        "       &   315.6       &  259.0        &     168.7         \\
NGC\,330    &    45 to 85    &  0.35 to 0.40  &    52.5       &   69.3        &      65.0         \\
NGC\,1761   &    60 to 120   &        "       &   571.8       &  537.8        &     463.0         \\
NGC\,1761   &   130 to 190   &        "       &    55.3       &   72.4        &      82.1         \\
NGC\,2004   &    40 to 100   &        "       &      --       &     --        &       --          \\   
NGC\,2004   &   100 to 180   &        "       &      --       &     --        &       --          \\   
NGC\,330    &    45 to 85    &  0.40 to 0.45  &    47.3       &   38.2        &      22.4         \\
NGC\,1761   &    60 to 120   &        "       &      --       &     --        &       --          \\
NGC\,1761   &   130 to 190   &        "       &      --       &     --        &       --          \\
NGC\,2004   &    40 to 100   &        "       &      --       &     --        &       --          \\
NGC\,2004   &   100 to 180   &        "       &      --       &     --        &       --          \\   
\hline
\end{tabular}
\end{center}
\end{table*}

\subsection{Search for molecular gas towards and in the LMC in FLAMES spectra}

Claims of molecular hydrogen detections in I/HVCs towards the LMC have been put forward by Richter et al. (1999) and Bluhm et al. (2001) 
using ORFEUS (Orbiting Retrievable Far and Extreme Ultraviolet Spectrometer) data. 
Additionally, Richter et al. (2003) obtained FUSE observations towards the LMC star Sk -68 80 and found IVC absorption 
at $\sim$+50 km\,s$^{-1}$ for 30 transitions and hints of molecular Hydrogen at HVC velocities of $\sim$+120 km\,s$^{-1}$, 
although being "too weak to claim a firm detection". Additionally, towards the LMC star Sk -68 82, Richter et al. (2003) co-added 15 H$_{2}$ 
transitions and found molecular gas at IVC and HVC velocities, although noting that the stellar continuum is very irregular which complicates 
the interpretation of the lines observed. Re-ananalysis of this star by Lehner et al. (2009) found no HVC detection in H$_{2}$. Likewise, 
although observations by Andr\'{e} et al. (2004) towards the Magellanic clouds using FUSE, HST and VLT observations have also shown the presence 
of H$_{2}$, HD and CO molecules, these detections were only at IVC, Milky Way or LMC/SMC velocities, with nothing seen corresponding to HVC gas. 
Searches for CO in emission towards HVCs have generally only given upper limits, (e.g. Dessauges-Zavadsky, Combes \& Pfenniger 2007), 
possibly due to the fact that small filaments of gas are unable to provide sufficient shielding from the ambient UV field (Richter et al. 2003). 
Overall, the presence of molecular gas in Magellanic HVCs is still a subject of debate. 

\begin{figure*}
\includegraphics[]{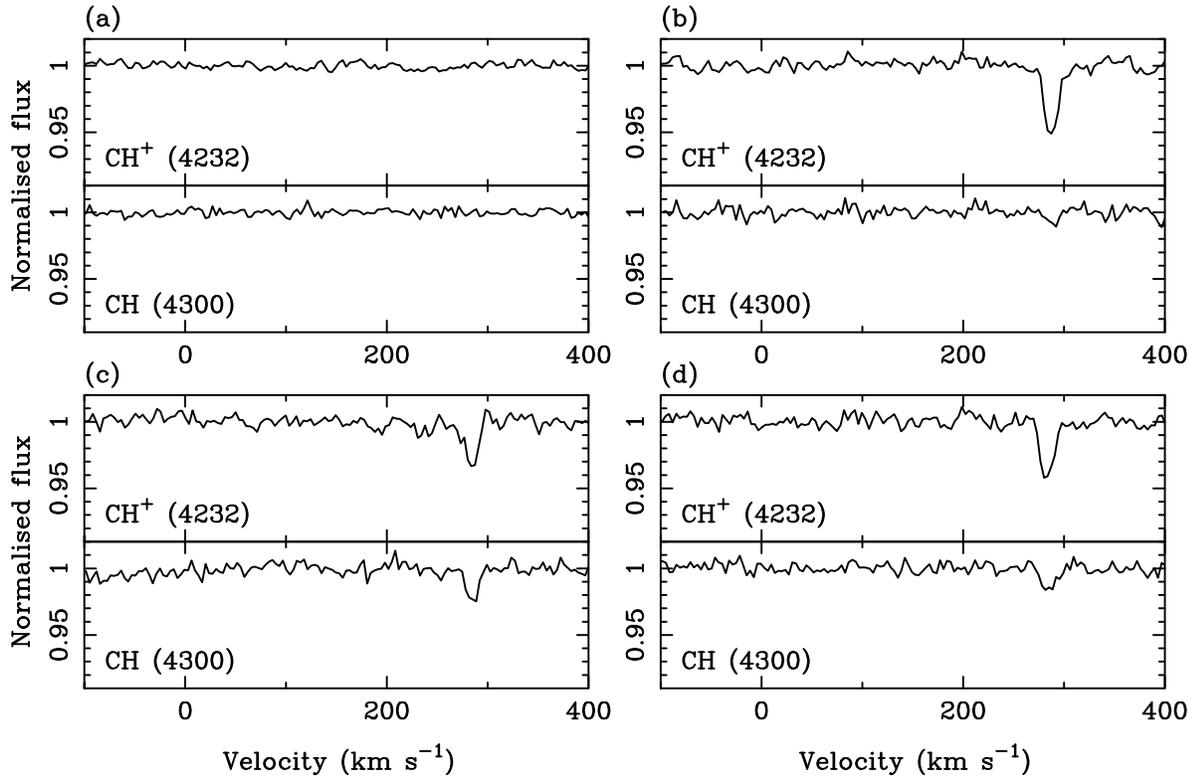}
\caption[]{CH$^{+}$ (4232\AA) and CH (4300\AA) profiles towards NGC\,1761 taken
with FLAMES-GIRAFFE. Only eight sightlines show molecular
line absorption at Magellanic Cloud velocities, and 
none show molecular absorption at Galactic or I/HVC velocities.}
\label{f_LMC_NGC_1761_Molecules}
\end{figure*}

Figures \ref{f_LMC_NGC_1761_Molecules} and \ref{f_LMC_NGC_2004_Molecules} show GIRAFFE spectra towards NGC\,1761 and NGC\,2004 in 
the molecular lines CH$^{+}$ (4232\AA) and CH (4300\AA). Tentative absorption in one or both of these species is detected in only three 
sightlines towards NGC\,1761 and one towards NGC\,2004 at the LMC velocity, with no absorption detected at Galactic 
or I/HVC velocities. We note that due to the relatively low gas density in HVCs, the detection of CH and CH$^{+}$ (although not H$_{2}$) 
is a-priori unlikely in equilibrium conditions. Na\,{\sc i} is rarely seen in absorption in HVCs, and CH is very well correlated with Na\,{\sc i} in 
the range log($N$(Na\,{\sc i}) cm$^{-2}$)$\sim$12.2-14.2 (Smoker et al. 2014).

The maximum equivalent widths measured in CH$^{+}$ (4232\AA) are 12.4 m\AA\,
for NGC\,1761 and 6.3 m\AA\, for NGC\,2004, respectively. For CH (4300\AA), the corresponding
values was 4.7 m\AA\, for NGC\,1761 with no obvious CH detection towards NGC\,2004. Towards NGC\,1761 a few 
of our sightlines have S/N ratios exceeding 400, which leads to a 3$\sigma$ detection limit of 1.7 m\AA \, or a 
EW variation exceeding $\sim$7 on scales of $\sim$10 arcminutes. 

Our detection rate is much lower than in the UVES spectra of Welty et al. (2006), who found either CH and/or CH$^{+}$ 
in  9 out of 13 LMC stars observed with UVES, likely due to the lower S/N ratio in many of our sightlines. 
However, the observed equivalent widths in the two samples are similar, with Welty et al. finding EWs from 0.5 to 13.0 m\AA\, for 
CH$^{+}$ (4232\AA) and 0.8 to 10.5 m\AA\, for CH (4300\AA). 

Finally we note that the absence of a strong molecular component in the LMC HVCs is consistent 
with the clouds being predominantly ionised and of a similar 
type to those studied by Lehner et al. (2009) using FUSE UV spectra. They found a lack of H$_{2}$ and a 
high ionisation level (average hydrogen ionization fraction $>$50\%), with $\approx$90\% of their sightlines also showing O\,{\sc vi}, 
indicating a diffuse, high-temperature component.

\begin{figure*}
\includegraphics[]{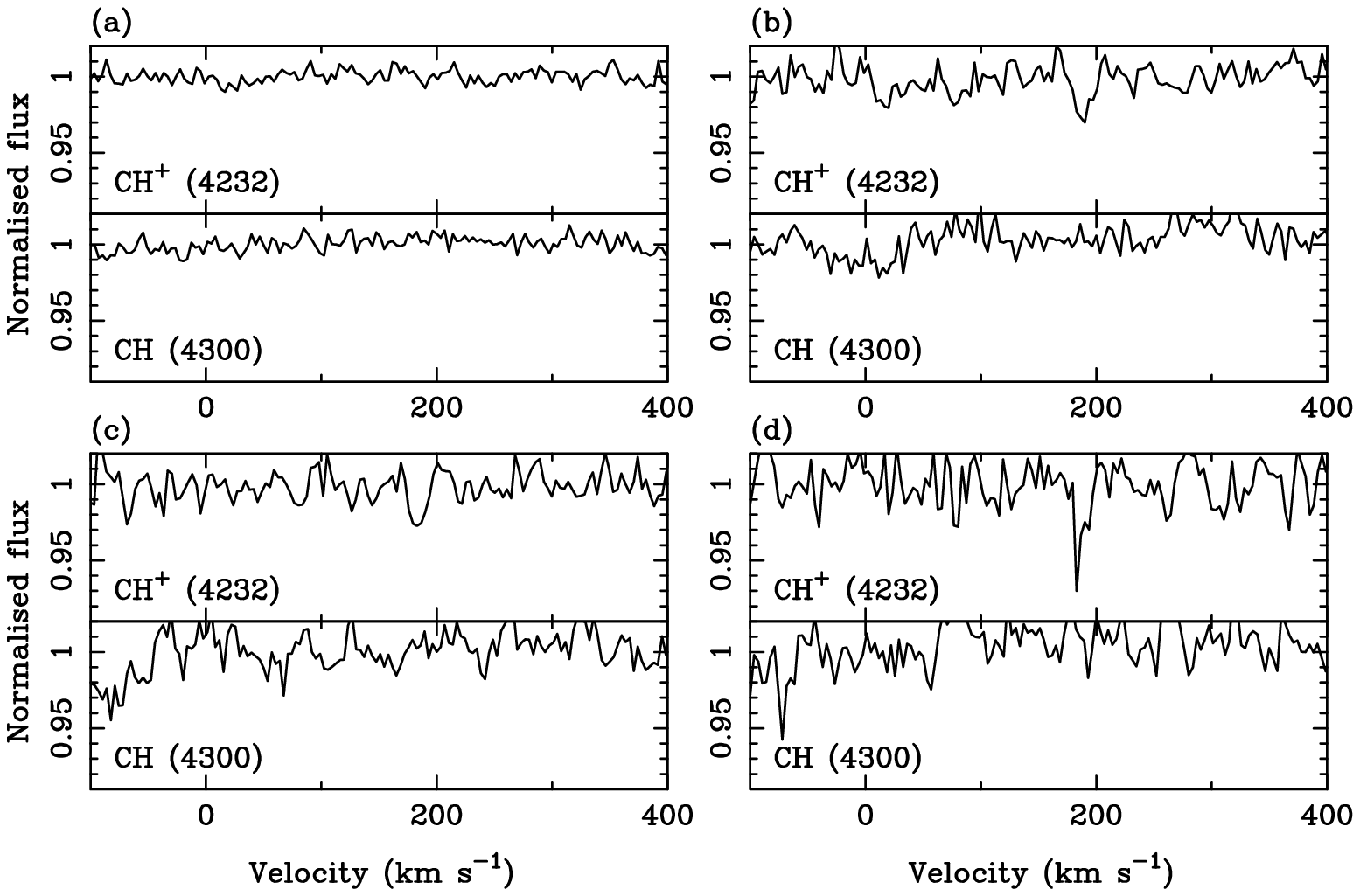}
\caption[]{CH$^{+}$ (4232\AA) and CH (4300\AA) profiles towards NGC\,2004. Only
three sightlines show tentative molecular line absorption at
Magellanic Cloud velocities, and none show molecular absorption at Galactic or I/HVC velocities.}
\label{f_LMC_NGC_2004_Molecules}
\end{figure*}

\section{Summary}
\label{Summary}

We have presented FEROS and FLAMES optical absorption line observations of intermediate and high velocity clouds toward
target stars within clusters in the LMC and SMC. IVC or HVC absorption in Ca\,{\sc ii} K is detected in $\sim$60 per
cent of the FEROS and UVES sightlines and in many 
of the LMC FLAMES-GIRAFFE sightlines. In the I/HVCs we find a variation in the observed Ca\,{\sc ii} equivalent 
width of a factor of $\sim$10 over $\sim$10 arcminutes or $\sim$150 pc at the distance of the LMC. 
Na\,{\sc i} D is only tentatively detected at high velocities in one sightline, indicating the Routly-Spitzer effect. 
The HV gas towards NGC\,2004 displays 50 to 100 percent more small-scale variation than the IV gas, indicating 
a structural difference between the two types of cloud. However, this is likely caused by the fact that 
the HV gas is further away and in any case the same difference is not seen towards our other LMC sightline NGC\,1761.

In the few sightlines with good H\,{\sc i} data the velocities of Ca\,{\sc ii} and H\,{\sc i} are the same within the errors, 
indicating that the two species are co-spatial. The Ca\,{\sc ii}/H\,{\sc i} ratios are higher in gas with velocities 
less than 70 km\,s$^{-1}$ than in HV gas, likewise the sightine with lowest Ca/O ratio is the one with the lowest 
velocity. These results are consistent with previous work indicating the the metalicities of IVCs tend to be closer 
to solar than for HVCs. Combining these Ca\,{\sc ii} observations with O\,{\sc i} measurements from 
the literature, we conclude that dust is present at HVC velocities in the LMC sightlines. 

Finally, we detect CH or CH$^{+}$ molecular gas in only four sightlines (and only at Magellanic 
velocities and not in the Milky Way or I/HVC components).



\section*{acknowledgements} \label{s_acknowledgments}

Data from ESO FEROS programme ID 078.C-0493(A) and FLAMES programme 171.D-0237(B) 
were taken from the ESO archive. This research has made use of the {\sc simbad} Database, 
operated at CDS, Strasbourg, France  and the ESO Archive. JVS thanks the ESO Director 
General Discretionary Fund and Queen's University Belfast Visiting Scientist Fund for 
financial support. We would like to thank two anonymous referees whose comments 
significantly improved the paper.


{}

\clearpage
\newpage

\appendix
\section{Online material}

\renewcommand\thefigure{\Alph{section}\arabic{figure}}
\renewcommand\thetable{\Alph{section}\arabic{table}}

\begin{table*}
\setcounter{table}{0}
\begin{center}
\caption[]{FEROS and UVES-observed stars towards the Magellanic Clouds. Stars with RA$<$2$^{\rm h}$ are 
in the SMC and with RA$>$3$^{\rm h}$ in the LMC. S/N ratios are per pixel.}
\label{t_FEROS_LMC_SMC}
\begin{tabular}{lrrrrrrrrrrr} 
\hline
Name          & Alt name & R.A.        & Dec.          & $l$   & $b$&m$_{v}$ &m$_{b}$   & Spect.        & S/N    & S/N   & Inst.  \\
              &          & (J2000)     & (J2000)       & (deg.)& (deg.) &(mag) & (mag) & type          & CaK    & NaD   &        \\
\hline

HV 1328       & SMC V0172    & 00 32 54.90  &  -73 49 19.1  &304.70&-43.24 &14.17  &14.116  &--             &    --  &  55   & U  \\ 
HV 1333       & SMC V0189    & 00 36 03.46  &  -73 55 58.8  &304.39&-43.15 &15.87  &14.702  &--             &    --  &  35   & U  \\ 
HV 1335       & SMC V0193    & 00 36 55.70  &  -73 56 27.9  &304.31&-43.15 &15.399 &14.746  &--             &    --  &  50   & U  \\ 
HV 817        & --           & 00 39 16.55  &  -72 01 58.4  &304.26&-45.06 &14.44  &13.77   &--             &    --  &  50   & U  \\ 
HV 1345       & SMC V0214    & 00 40 38.60  &  -73 13 14.2  &304.01&-43.88 &15.428 &14.779  &--             &    --  &  15   & U  \\ 
HV 822        & --           & 00 41 55.48  &  -73 32 23.7  &303.86&-43.57 &14.39  &13.95   &--             &    --  &  30   & U  \\ 
HV 1365       & --           & 00 42 00.00  &  -73 44 00.0  &303.84&-43.38 &15.55  &14.94   &--             &    --  &  30   & U  \\ 
SV* HV 823    & --           & 00 43 48.50  &  -73 36 50.0  &303.67&-43.50 &13.99  &13.39   &--             &    --  &  60   & U  \\ 
AzV 014       & Sk 9         & 00 46 32.63  &  -73 06 05.7  &303.43&-44.02 &13.59  &13.77   &O3-4V+neb      &    60  &  --   & U  \\ 
AzV 015       & Sk 10        & 00 46 42.14  &  -73 24 55.7  &303.40&-43.71 &12.984 &13.176  &O6.5II(f)      &    60  &  --   & U  \\ 
AzV 018       & Sk 13        & 00 47 12.22  &  -73 06 33.2  &303.36&-44.01 &12.52  &12.48   &OB             &   120  & 100   & U  \\ 
AzV 022       & Sk 15        & 00 47 38.70  &  -73 07 48.8  &303.31&-44.00 &12.15  &12.25   &B2I            &    70  &  --   & U  \\ 
AzV 023       & Sk 17        & 00 47 38.91  &  -73 22 53.9  &303.31&-43.74 &12.324 &12.236  &B2I            &   120  & 200   & U  \\ 
SMC 9251      & --           & 00 47 47.60  &  -73 17 28.0  &303.30&-43.83 &14.90  &14.76   &--             &    70  &  --   & U  \\ 
AzV 026       & Sk 18        & 00 47 50.04  &  -73 08 21.0  &303.30&-43.99 &12.35  &12.55   &O7III+neb      &   250  &  --   & U  \\ 
AzV 047       & SMC 11925    & 00 48 51.48  &  -73 25 58.5  &303.19&-43.69 &13.314 &13.496  &O8III((f))     &    60  &  60   & U  \\ 
AzV 065       & Sk 33        & 00 50 06.08  &  -73 07 45.2  &303.07&-44.00 &11.12  &11.00   &B5I            &   100  & 200   & U  \\ 
AzV 070       & Sk 35        & 00 50 18.11  &  -72 38 10.0  &303.05&-44.49 &10.96  &12.40   &O9.5Iw         &   130  & 130   & U  \\ 
AzV 075       & Sk 38        & 00 50 32.39  &  -72 52 36.5  &303.02&-44.25 &12.604 &12.756  &O5III(f+)      &    75  &  --   & U  \\ 
AzV 80        & SMC 17457    & 00 50 43.80  &  -72 47 41.5  &303.01&-44.33 &13.24  &13.376  &O4-6n(f)p      &   100  &  80   & U  \\ 
DZ Tuc        & SMC 17504    & 00 50 44.70  &  -73 16 05.4  &303.00&-43.86 &15.37  &15.44   &B0-B0.5V       &    25  &  30   & U  \\ 
AzV 95        & SMC 19650    & 00 51 21.65  &  -72 44 14.4  &302.94&-44.39 &13.64  &13.83   &O7III((f))     &   100  &  60   & U  \\ 
AzV 104       & SMC 20656    & 00 51 38.43  &  -72 48 06.1  &302.91&-44.33 &13.064 &13.226  &B0.5Ia         &    70  &  --   & U  \\ 
LIN 200       & Bruck 60 14  & 00 51 57.74  &  -73 14 22.0  &302.88&-43.89 &17.493 &17.20   &--             &    10  &  50   & U  \\ 
AzV 120       & SMC 23151    & 00 52 15.24  &  -72 09 15.8  &302.84&-44.97 &14.14  &14.43   &O9.5III        &   120  &  60   & U  \\ 
ESHC 05       & --           & 00 52 06.43  &  -73 06 29.4  &302.86&-44.02 &14.2   &13.983  &--             &    --  & 100   & U  \\ 
ESHC 03       & --           & 00 52 21.50  &  -73 13 33.0  &302.84&-43.90 &    -- &    --  &--             &    30  &  25   & U  \\ 
ESHC 02       & --           & 00 52 32.60  &  -73 17 08.0  &302.82&-43.84 &16.957 &17.012  &B2IV-V         &    20  &  35   & U  \\ 
ESHC 07       & --           & 00 52 52.60  &  -73 18 34.0  &302.79&-43.82 &15.417 &15.254  &--             &   120  & 250   & U  \\ 
ESHC 01       & LIN 232      & 00 53 02.80  &  -73 17 59.4  &302.77&-43.83 &15.072 &15.079  &B4III          &    20  &  --   & U  \\ 
ESHC 04       & --           & 00 53 56.75  &  -73 10 29.4  &302.68&-43.95 &14.901 &15.049  &--             &    40  &  40   & U  \\ 
OGLE 13487    & --           & 00 54 33.22  &  -73 10 39.0  &302.62&-43.95 &15.51  &15.69   &--             &   200  &  --   & U  \\ 
ESHC 06       & LIN 264      & 00 54 37.60  &  -73 04 56.0  &302.61&-44.04 &15.128 &15.177  &--             &    60  &  30   & U  \\ 
HV 1645       & --           & 00 55 19.76  &  -73 14 42.2  &302.54&-43.88 &20.10  &16.84   &Mira           &    --  &  20   & U  \\ 
LHA 115-S 23  & AzV 172      & 00 55 53.81  &  -72 08 59.0  &302.45&-44.97 &13.29  &13.25   &B8[e]Ib        &    10  &  30   & F  \\ 
N330 ROB B18  & SMC 35727    & 00 56 03.70  &  -72 27 12.8  &302.44&-44.67 &15.57  &15.79   &O9.5V          &    30  &  --   & U  \\ 
N330 ROB B28  & N330 BAL 254 & 00 56 06.80  &  -72 28 34.9  &302.44&-44.65 &15.52  &15.71   &B0Ve           &    30  &  --   & U  \\ 
N330 ROB B32  & N330 ELS 36  & 00 56 10.65  &  -72 28 10.1  &302.43&-44.65 &14.74  &14.85   &B2II           &    35  &  --   & U  \\ 
N330 ROB B16  & --           & 00 56 18.56  &  -72 26 45.4  &302.42&-44.68 &13.92  &13.98   &A2II           &    60  &  40   & U  \\ 
N330 ROB B13  & N330 ELS 125 & 00 56 20.11  &  -72 27 02.3  &302.41&-44.67 &15.58  &15.78   &B2III/IVe      &    20  &  --   & U  \\ 
N330 ROB B38  & N330 ELS 121 & 00 56 22.56  &  -72 28 35.9  &302.41&-44.65 &12.474 &12.366  &A5I            &    --  &  70   & U  \\ 
N330 ROB B04  & --           & 00 56 30.90  &  -72 28 19.0  &302.39&-44.65 &15.41  &15.58   &B1.5IVe        &    25  &  --   & U  \\ 
AzV 187       & Sk 68        & 00 57 31.73  &  -71 19 59.3  &302.23&-45.79 &11.96  &12.09   &OB             &    55  &  85   & F  \\ 
N346 ELS 26   & SMC 42719    & 00 58 14.24  &  -72 10 45.0  &302.20&-44.94 &19.15  &17.98   &B0IV(Nstr)     &    --  &  65   & U  \\ 
N346 ELS 12   & AzV 202      & 00 58 14.48  &  -72 07 29.8  &302.19&-44.99 &14.18  &14.34   &B1Ib           &    --  &  60   & U  \\ 
N346 ELS 28   & --           & 00 58 17.35  &  -72 10 50.8  &302.19&-44.94 &14.73  &14.94   &OC6Vz          &    --  &  70   & U  \\ 
AzV 207       & SMC 43724    & 00 58 33.17  &  -71 55 47.0  &302.15&-45.19 &14.13  &14.35   &O7V            &    40  &  60   & U  \\ 
AzV 210       & Sk 73        & 00 58 35.79  &  -72 16 25.0  &302.16&-44.84 &12.63  &12.66   &OB             &    50  &  --   & U  \\ 
N346 ELS 18   & N346 ELS 18  & 00 58 47.03  &  -72 13 01.6  &302.14&-44.90 &14.66  &14.78   &O9.5IIIe       &    --  &  40   & U  \\ 
AzV 214       & SMC 44784    & 00 58 54.76  &  -72 13 17.2  &302.13&-44.90 &13.40  &13.39   &B1Ia           &   100  &  90   & U  \\ 
AzV 215       & Sk 76        & 00 58 55.62  &  -72 32 08.5  &302.14&-44.58 &12.65  &12.75   &OB             &    60  &  --   & U  \\ 
N346 ELS 07   & SMC 44908    & 00 58 57.40  &  -72 10 33.5  &302.12&-44.94 &13.82  &14.13   &O4V((f))       &    --  & 110   & U  \\ 
AzV 216       & --           & 00 58 59.13  &  -72 44 34.0  &302.15&-44.37 &14.15  &14.32   &B1-3II:        &    40  &  --   & U  \\ 
N346 MPG 355  & SMC 45068    & 00 59 00.75  &  -72 10 28.2  &302.11&-44.94 &12.61  &12.80   &O2III(f)       &    --  & 100   & U  \\ 
N346 MPG 368  & --           & 00 59 01.80  &  -72 10 31.2  &302.11&-44.94 &13.95  &14.18   &O6:V           &    --  & 120   & U  \\ 
N346 MPG 487  & --           & 00 59 06.71  &  -72 10 41.3  &302.10&-44.94 &14.19  &14.33   &O8V            &    --  &  70   & U  \\ 
\hline
\end{tabular}
\end{center}
\end{table*}

\begin{table*}
\setcounter{table}{0}
\begin{center}
\caption[]{continued.}
\label{t_FEROS_LMC_SMC}
\begin{tabular}{lrrrrrrrrrrr} 
\hline
Name          & Alt name & R.A.        & Dec.          & $l$   & $b$&m$_{v}$ &m$_{b}$   & Spect.        & S/N    & S/N   & Inst.  \\
              &          & (J2000)     & (J2000)       & (deg.)& (deg.) &(mag) & (mag) & type          & CaK    & NaD   &        \\
\hline
N346 ELS 51   & SMC 45459    & 00 59 08.68  &  -72 10 14.1  &302.10&-44.95 &15.17  &15.40   &O7Vz           &    --  &  30   & U  \\ 
N346 ELS 33   & N346 SSN 37  & 00 59 11.64  &  -72 09 57.6  &302.09&-44.95 &14.82  &15.07   &O8V            &    --  &  30   & U  \\ 
N346 ELS 22   & SMC 45935    & 00 59 18.58  &  -72 11 10.1  &302.08&-44.93 &14.65  &14.91   &O9V            &    --  &  40   & U  \\ 
N346 ELS 10   & AzV 226      & 00 59 20.73  &  -72 17 10.7  &302.08&-44.83 &14.12  &14.19   &O7IIIn((f))    &    --  &  80   & U  \\ 
HD 5980       & AzV 229      & 00 59 26.57  &  -72 09 53.9  &302.07&-44.95 &11.13  &11.31   &WN3+OB         &    --  &  60   & U  \\ 
N346 ELS 46   & --           & 00 59 31.88  &  -72 13 35.2  &302.06&-44.89 &15.16  &15.44   &O7Vn           &    --  &  40   & U  \\ 
AzV 232       & Sk 80        & 00 59 31.96  &  -72 10 46.3  &302.06&-44.93 &12.12  &12.31   &O7Iaf+         &   120  & 150   & U  \\ 
AzV 235       & Sk 82        & 00 59 45.75  &  -72 44 56.5  &302.07&-44.37 &12.02  &12.20   &B0Iaw          &   100  &  --   & U  \\ 
N346 ELS 31   & SMC 47478    & 00 59 54.08  &  -72 04 31.0  &302.01&-45.04 &14.81  &15.05   &O8Vz           &    --  &  60   & U  \\ 
AzV 243       & Sk 84        & 01 00 06.71  &  -72 47 19.0  &302.03&-44.32 &13.71  &13.91   &O6V            &    80  &  --   & U  \\ 
AzV 242       & Sk 85        & 01 00 06.86  &  -72 13 57.5  &301.10&-44.88 &11.7   &11.5    &B0.7Iaw        &   250  & 200   & U  \\ 
AzV 304       & SMC 53474    & 01 02 21.47  &  -72 39 14.7  &301.79&-44.45 &14.66  &14.77   &B0.5V          &    70  &  --   & U  \\ 
AzV 321       & SMC 54958    & 01 02 57.07  &  -72 08 09.1  &301.68&-44.96 &13.66  &13.82   &O9Ib           &   110  &  70   & U  \\ 
AzV 372       & Sk 116       & 01 04 55.74  &  -72 46 48.1  &301.54&-44.31 &12.494 &12.646  &O9.5Iabw       &    90  &  --   & U  \\ 
AzV 388       & SMC 62400    & 01 05 39.52  &  -72 29 27.1  &301.43&-44.60 &13.86  &14.12   &O4V            &    50  &  --   & U  \\ 
AzV 398       & SMC 63413    & 01 06 09.81  &  -71 56 00.8  &301.31&-45.15 &13.94  &13.98   &O9Ia:          &    80  &  65   & U  \\ 
MA93 1589     & --           & 01 06 28.85  &  -71 52 04.9  &301.27&-45.21 &15.411 &15.135  &A5             &   100  &  --   & U  \\ 
AzV 404       & Sk 128       & 01 06 29.27  &  -72 22 08.4  &301.33&-44.71 &11.7   &11.8    &OB             &   200  & 120   & U  \\ 
LHA 115-S 52  & HD 6884      & 01 07 18.22  &  -72 28 03.7  &301.25&-44.61 &10.32  &10.23   &B9Iae          &    45  & 125   & F  \\ 
AzV 440       & SMC 68756    & 01 08 56.01  &  -71 52 46.8  &301.00&-45.18 &14.44  &14.58   &O7V            &    60  &  35   & U  \\ 
AzV 456       & Sk 143       & 01 10 55.77  &  -72 42 56.3  &300.91&-44.34 &12.98  &12.89   &OB             &    70  & 200   & U  \\ 
AzV 462       & Sk 145       & 01 11 25.92  &  -72 31 20.9  &300.83&-44.52 &12.464 &12.596  &OB             &    30  &  65   & F  \\ 
AzV 469       & Sk 148       & 01 12 29.00  &  -72 29 29.1  &300.71&-44.55 &13.014 &13.176  &O8II           &   100  &  --   & U  \\ 
HV 2195       & --           & 01 14 28.05  &  -72 39 53.5  &300.54&-44.36 &12.99  &12.51   &F5Ib           &    --  &  50   & U  \\ 
AzV 488       & Sk 159       & 01 15 08.88  &  -73 21 24.3  &300.51&-43.66 &11.77  &11.90   &B0.5Iaw        &   130  & 130   & U  \\ 
AzV 483       & Sk 156       & 01 15 28.63  &  -73 19 50.1  &300.55&-43.69 &11.85  &11.93   &OB             &    50  &  70   & F  \\ 
SK 160        & AzV 190      & 01 17 05.15  &  -73 26 36.0  &300.41&-43.56 &13.12  &13.30   &OB:            &   120  & 150   & U  \\ 
SK 190        & --           & 01 31 27.96  &  -73 22 14.3  &299.00&-43.45 &13.37  &13.59   &O7.5(f)np      &    90  &  65   & U  \\ 
SK-67 2       & RMC 51       & 04 47 04.45  &  -67 06 53.1  &278.36&-36.79 &11.219 &11.26   &B1.5Ia         &    50  & 130   & F  \\ 
SK -67 05     & RMC 53       & 04 50 18.92  &  -67 39 38.1  &278.89&-36.32 &11.2   &11.377  &O9.7Ib         &   220  &  --   & U  \\ 
SK-66 1       & RMC 56       & 04 52 19.09  &  -66 43 53.3  &277.72&-36.41 &11.603 &11.635  &B2Ia           &    30  &  60   & F  \\ 
SK-66 5       & RMC 57       & 04 53 30.03  &  -66 55 28.3  &277.90&-36.24 &10.68  &10.76   &B3Iab          &    30  & 130   & F  \\ 
SK-67 14      & Sk -67 14    & 04 54 31.89  &  -67 15 24.7  &278.27&-36.05 &11.376 &11.541  &B1.5Ia         &    40  &  80   & F  \\ 
SK -66 18     & --           & 04 55 59.80  &  -65 58 29.8  &276.69&-36.25 &13.294 &13.467  &O6V((f))       &   100  &  80   & U  \\ 
SK -69 43     & HD 268809    & 04 56 10.46  &  -69 15 38.2  &280.58&-35.33 &11.866 &11.964  &OB             &    30  &  70   & F  \\ 
LH 10-3061    & --           & 04 56 42.46  &  -66 25 18.1  &277.20&-36.07 &13.595 &13.491  &ON2III(f*)     &    90  &  75   & U  \\ 
SK -66 35     & HD 268732    & 04 57 04.47  &  -66 34 38.5  &277.38&-35.99 &11.494 &11.587  &OB             &    50  &  80   & F  \\ 
SK -69 50     & LMC 33053    & 04 57 15.09  &  -69 20 19.9  &280.64&-35.22 &13.204 &13.366  &O7(f)(n)p      &   100  &  80   & U  \\ 
SK -67 22     & Brey 10a     & 04 57 27.44  &  -67 39 02.9  &278.64&-35.67 &13.314 &13.496  &Of             &   110  &  --   & U  \\ 
LHA 120-S 12  & SK -67 23    & 04 57 36.80  &  -67 47 37.5  &278.80&-35.62 &12.66  &12.526  &B0.5Ie         &    35  &  70   & F  \\ 
SK -69 52     & HD 268867    & 04 57 48.90  &  -69 52 22.3  &281.24&-35.02 &11.392 &11.399  &OB             &    50  &  --   & U  \\ 
SK -67 28     & GV 135       & 04 58 39.24  &  -67 11 18.7  &278.05&-35.69 &12.082 &12.269  &B0.7Ia         &    35  &  85   & F  \\ 
SK -68 26     & GV 167       & 05 01 32.24  &  -68 10 42.9  &279.14&-35.17 &11.728 &11.85   &BC2Ia          &    35  &  70   & F  \\ 
LHA 120-S 155 & RMC 71       & 05 02 07.39  &  -71 20 13.1  &282.82&-34.25 &10.60  &10.55   &OB             &    35  &  70   & F  \\ 
SK -69 59     & HD 268960    & 05 03 12.70  &  -69 01 37.0  &280.09&-34.80 &12.01  &12.166  &B0Ia           &    25  &  45   & F  \\ 
SK -67 38     & LMC 59721    & 05 03 29.73  &  -67 52 25.1  &278.72&-35.07 &13.494 &13.716  &OB             &   100  &  80   & U  \\ 
SK -70 50     & HD 269009    & 05 03 45.85  &  -70 11 57.5  &281.45&-34.44 &11.0   &11.1    &OB             &    50  & 105   & F  \\ 
SK -70 60     & LMC 64006    & 05 04 40.78  &  -70 15 34.6  &281.49&-34.35 &13.646 &13.914  &O4-O5V:n       &   110  &  --   & U  \\ 
SK -68 39     & --           & 05 04 50.17  &  -68 07 52.4  &278.99&-34.88 &11.988 &12.039  &B2.5Ia         &    50  & 115   & F  \\ 
SK -70 69     & LMC 65981    & 05 05 18.70  &  -70 25 49.8  &281.68&-34.25 &13.616 &13.854  &O3V(f)         &    75  &  50   & U  \\ 
SK -68 41     & GV 195       & 05 05 27.11  &  -68 10 02.6  &279.02&-34.82 &11.868 &12.01   &B0.5Ia         &    40  & 100   & F  \\ 
SK -68 45     & --           & 05 06 07.28  &  -68 07 06.2  &278.94&-34.77 &11.92  &12.007  &OB             &    30  &  70   & F  \\ 
SK -70 78     & HD 269074    & 05 06 16.04  &  -70 29 35.7  &281.72&-34.16 &11.0   &11.1    &OB             &    50  &  70   & F  \\ 
SK -68 52     & HD 269050    & 05 07 20.41  &  -68 32 08.6  &279.40&-34.56 &11.581 &11.54   &B0Ia           &    50  &  --   & U  \\ 
OGLE J050724  & --           & 05 07 24.62  &  -68 29 32.7  &279.35&-34.56 &   --  &16.76   &Ecl. Bin       &    45  &  --   & U  \\ 
SK -68 63     & HIP 24080    & 05 10 22.79  &  -68 46 23.8  &279.60&-34.23 &10.52  &10.52   &B1.5eq         &   200  &  --   & U  \\ 
\hline
\end{tabular}
\end{center}
\end{table*}

\begin{table*}
\setcounter{table}{0}
\begin{center}
\caption[]{continued.}
\begin{tabular}{lrrrrrrrrrrr} 
\hline
Name          & Alt name & R.A.        & Dec.          & $l$   & $b$&m$_{v}$ &m$_{b}$   & Spect.        & S/N    & S/N   & Inst.  \\
              &          & (J2000)     & (J2000)       & (deg.)& (deg.) &(mag) & (mag) & type          & CaK    & NaD   &        \\
\hline
BI 108        & LMC 94226    & 05 13 43.05  &  -69 18 36.9  &280.15&-33.82 &13.222 &13.332  &B1:II:         &   130  &  --   & U  \\ 
SK -69 83     & HD 269244    & 05 14 29.64  &  -69 29 43.4  &280.35&-33.72 &11.435 &11.612  &OB             &    30  &  70   & F  \\ 
LHA 120-S 93  & GV 566       & 05 16 31.80  &  -68 22 09.1  &278.98&-33.77 &12.81  &12.69   &A0:I:          &    25  &  45   & F  \\ 
SK -69 89     & HD  269311   & 05 17 17.57  &  -69 46 44.2  &280.62&-33.42 &11.999 &11.428  &OB             &    40  &  80   & F  \\ 
LHA 120-S 96  & SK -69 94    & 05 18 14.35  &  -69 15 01.1  &279.98&-33.44 & 9.635 & 9.565  &A5Iaeq         &    50  & 165   & F  \\ 
BI 128        & --           & 05 18 19.77  &  -65 49 14.6  &275.92&-34.06 &13.447 &13.745  &O...           &    80  &  40   & U  \\ 
SK -67 78     & HD 269371    & 05 20 19.08  &  -67 18 05.7  &277.64&-33.62 &11.3   &11.0    &OB             &    30  &  95   & F  \\ 
LHA 120-S 30  & SK -68 73    & 05 22 59.73  &  -68 01 46.3  &278.45&-33.24 &11.71  &11.46   &Bep            &    35  &  95   & F  \\ 
SK -67 90     & HD 269440    & 05 23 00.66  &  -67 11 22.1  &277.46&-33.38 &11.282 &11.378  &B1Ia           &    55  & 120   & F  \\ 
SK -67 112    & HD 269545    & 05 26 56.48  &  -67 39 35.0  &277.94&-32.93 &11.77  &11.90   &OB             &    30  &  80   & F  \\ 
SK -66 100    & --           & 05 27 45.47  &  -66 55 15.2  &277.06&-32.96 &12.976 &13.204  &O6II(f)        &   110  &  70   & U  \\ 
SK -68 92     & GV 315       & 05 28 16.17  &  -68 51 45.6  &279.33&-32.63 &11.64  &11.71   &OB             &    45  &  80   & F  \\ 
HD 269599     & RMC 105      & 05 28 22.73  &  -69 08 31.7  &279.66&-32.60 &10.24  &10.03   &--             &    35  &  30   & F  \\ 
SK -66 106    & --           & 05 29 00.99  &  -66 38 27.8  &276.71&-32.87 &12.64  &12.72   &OB             &    55  & 120   & F  \\ 
LHA 120-S 116 & RMC 110      & 05 30 51.48  &  -69 02 58.6  &279.51&-32.37 &10.52  &10.28   &F0Iae          &    30  &  70   & F  \\ 
SK -66 118    & GV 341       & 05 30 51.91  &  -66 54 09.1  &276.99&-32.66 &11.674 &11.776  &OB             &    45  &  55   & F  \\ 
SK -67 150    & --           & 05 30 01.71  &  -67 00 53.4  &277.13&-32.68 &12.034 &12.24   &OB             &    50  & 110   & F  \\ 
SK -71 42     & HD 269660    & 05 30 47.78  &  -71 04 02.3  &281.87&-32.06 &11.15  &11.19   &B1Ia           &    55  &  95   & F  \\ 
SK -68 111    & HD 269668    & 05 31 00.84  &  -68 53 57.1  &279.33&-32.38 &10.161 &12.01   &OB             &    45  &  75   & F  \\ 
SK -67 169    & --           & 05 31 51.59  &  -67 02 22.3  &277.14&-32.55 &12.06  &12.18   &B1Ia           &    45  &  90   & F  \\ 
HDE 269702    & SK -67 168   & 05 31 52.12  &  -67 34 20.8  &277.76&-32.48 &11.69  &11.99   &O8I(f)p        &   125  &  70   & U  \\ 
SK -68 114    & HD 269700    & 05 31 52.28  &  -68 32 38.9  &278.90&-32.35 &10.56  &10.54   &BN?2Ia+        &    25  &  45   & F  \\ 
SK -67 172    & HD 269713    & 05 32 07.32  &  -67 29 14.0  &277.66&-32.47 &11.81  &11.88   &OB             &    30  &  70   & F  \\ 
SK -67 173    & LH 76-51     & 05 32 10.75  &  -67 40 25.0  &277.88&-32.44 &11.92  &12.04   &OB             &    30  &  45   & F  \\ 
SK -67 199    & HD 269777    & 05 34 18.45  &  -67 18 13.7  &277.41&-32.28 &11.05  &11.06   &B3Ia           &    40  &  80   & F  \\ 
SK -67 206    & --           & 05 34 55.11  &  -67 02 37.5  &277.10&-32.25 & 9.862 &12.00   &OB             &    45  &  80   & F  \\ 
BI 237        & LMC 164942   & 05 36 14.63  &  -67 39 19.2  &277.80&-32.06 &13.790 &13.830  &O2V(f*)        &    90  &  60   & U  \\ 
SK -69 214    & CPD-69 419   & 05 36 16.42  &  -69 31 27.0  &279.99&-31.83 &12.279 &12.19   &B0.7Ia         &    30  &  70   & F  \\ 
HD 37974      & SK -69 216   & 05 36 25.87  &  -69 22 55.9  &279.82&-31.84 &11.10  &10.959  &Be             &    25  &  90   & F  \\ 
SK -66 171    & HD 269889    & 05 37 02.42  &  -66 38 36.9  &276.61&-32.08 &10.246 &12.19   &OB             &    95  &  50   & U  \\ 
SK -69 228    & CPD-69 436   & 05 37 09.18  &  -69 20 19.5  &279.76&-31.78 &12.19  &12.12   &BC1.5Ia        &    40  & 105   & F  \\ 
BI 253        & LMC 168644   & 05 37 34.46  &  -69 01 10.2  &279.38&-31.78 &13.56  &13.70   &O2V(f*)        &   105  &  90   & U  \\ 
SK -69 237    & SK -69 237   & 05 38 01.31  &  -69 22 14.1  &279.79&-31.70 &12.05  &12.08   &B1Ia           &    30  &  90   & F  \\ 
LHA 120-S 134 & HD 38489     & 05 40 13.33  &  -69 22 46.5  &279.79&-31.70 &12.296 &12.215  &B[e]           &    30  &  70   & F  \\ 
SK-69 270     & HD 269997    & 05 41 21.19  &  -69 04 38.6  &279.41&-31.44 &11.36  &11.20   &B2.5Ia         &    40  & 110   & F  \\ 
SK-69 274     & HD 269992    & 05 41 27.68  &  -69 48 03.7  &280.25&-31.35 &11.27  &11.22   &B2.5Ia         &    45  & 120   & F  \\ 
SK-70 111     & HD 269993    & 05 41 36.79  &  -70 00 52.6  &280.50&-31.32 &11.78  &11.516  &OB             &    35  &  80   & F  \\ 
LHA 120-S 137 & --           & 05 41 48.00  &  -69 37 00.0  &280.03&-31.34 &14.198 &14.107  &BIaePCyg       &     5  &  15   & F  \\ 
SK-69 289     & CPD-69 514   & 05 42 49.22  &  -69 32 52.2  &279.94&-31.26 &11.4   &11.3    &OB             &    35  &  65   & F  \\ 
SK-67 256     & CPD-67 500   & 05 44 25.02  &  -67 13 49.6  &277.22&-31.32 &11.76  &11.751  &OB             &    20  &  85   & F  \\ 
SK-68 171     & HD 270220    & 05 50 22.99  &  -68 11 24.7  &278.30&-30.69 &11.960 &12.001  &B1Ia           &    20  &  55   & F  \\ 
SK-70 120     & HD 270196    & 05 51 20.78  &  -70 17 09.3  &280.72&-30.47 &11.490 &11.6    &B1Ia           &    35  &  60   & F  \\ 
\hline
\end{tabular}
\end{center}
\end{table*}

\clearpage
\newpage

\clearpage
\newpage

%
%

\begin{figure*}
\setcounter{figure}{0}
\includegraphics[height=22cm]{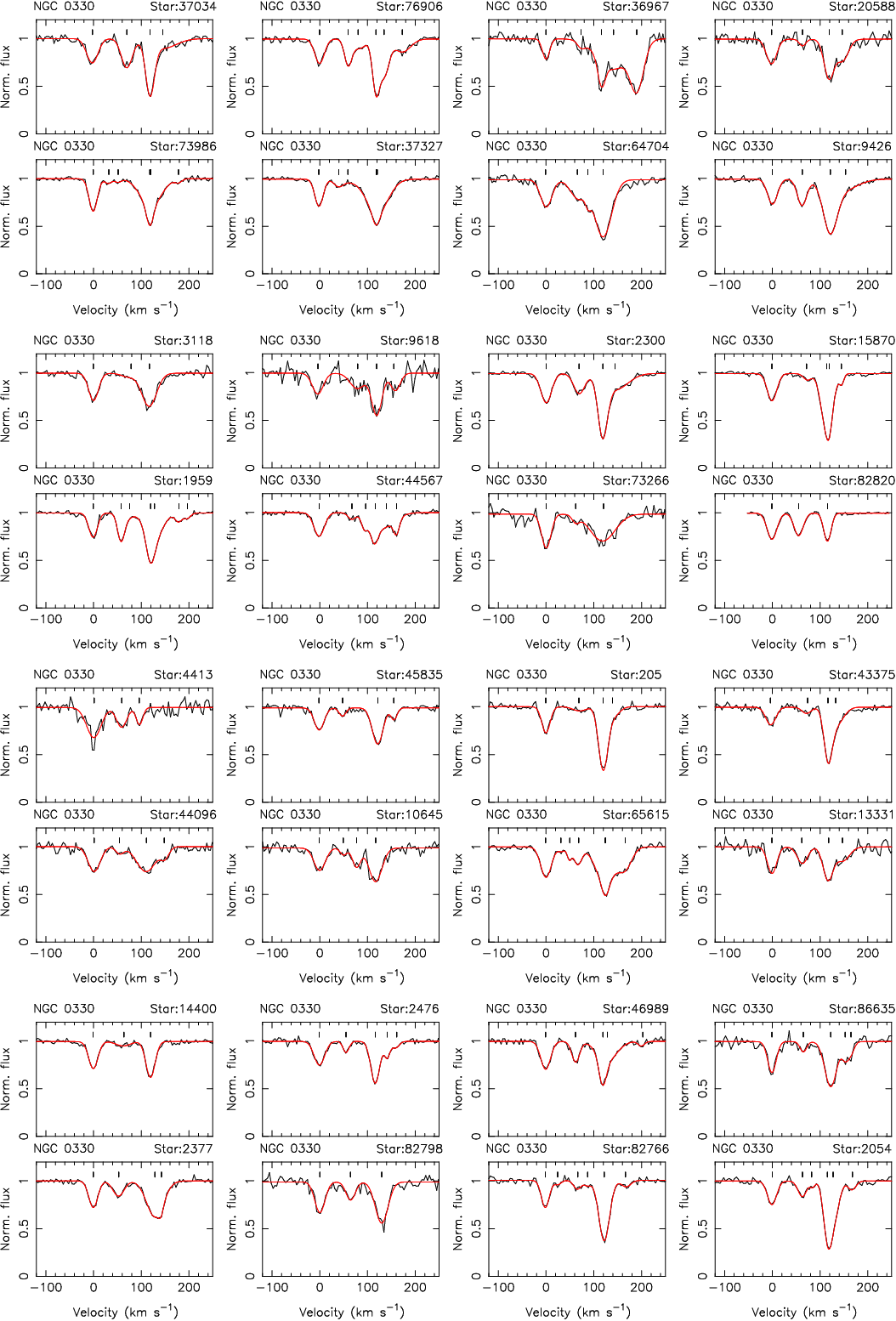}
\caption[]{FLAMES-GIRAFFE Ca\,{\sc ii} K spectra towards NGC\,330 showing the spectra of 16 star-to-star 
pairs in which the maximum difference in the equivalent width was detected 
between a velocity of +45 and +85 km\,s$^{-1}$ in the LSR. Each star is only plotted once.}
\label{f_MaxVariation_EW_I/HVCs_NGC_330_I/HVC_p45_p85}
\end{figure*}

\begin{figure*}
\setcounter{figure}{1}
\includegraphics[height=22cm]{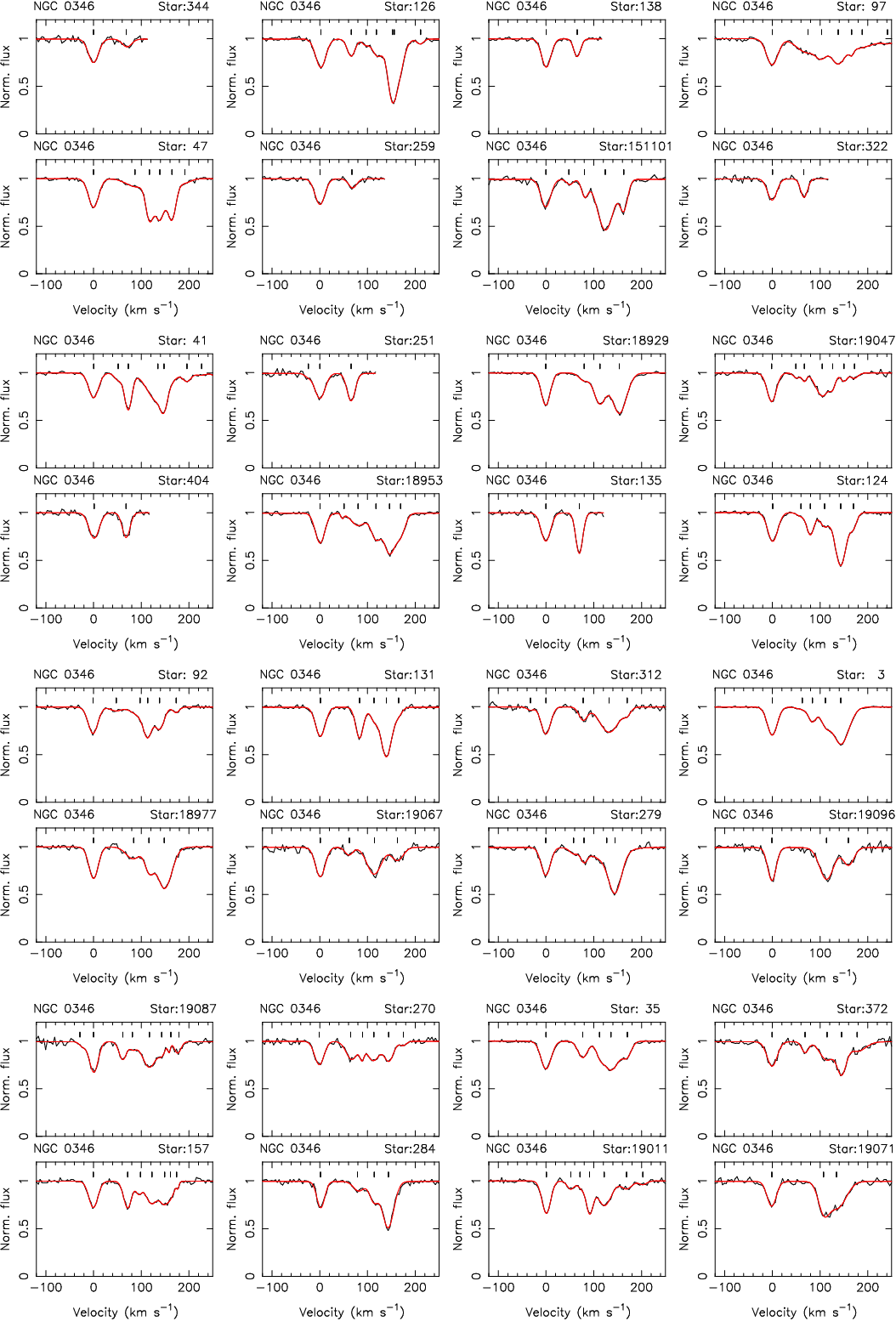}
\caption[]{FLAMES-GIRAFFE Ca\,{\sc ii} K spectra towards NGC\,346 showing the spectra of 16 star-to-star 
pairs in which the maximum difference in the equivalent width was detected 
between a velocity of +60 and +250 km\,s$^{-1}$ in the LSR. Each star is only plotted once.}
\label{f_MaxVariation_EW_I/HVCs_NGC_346_I/HVC_p60_p250}
\end{figure*}

\begin{figure*}
\setcounter{figure}{2}
\includegraphics[height=22cm]{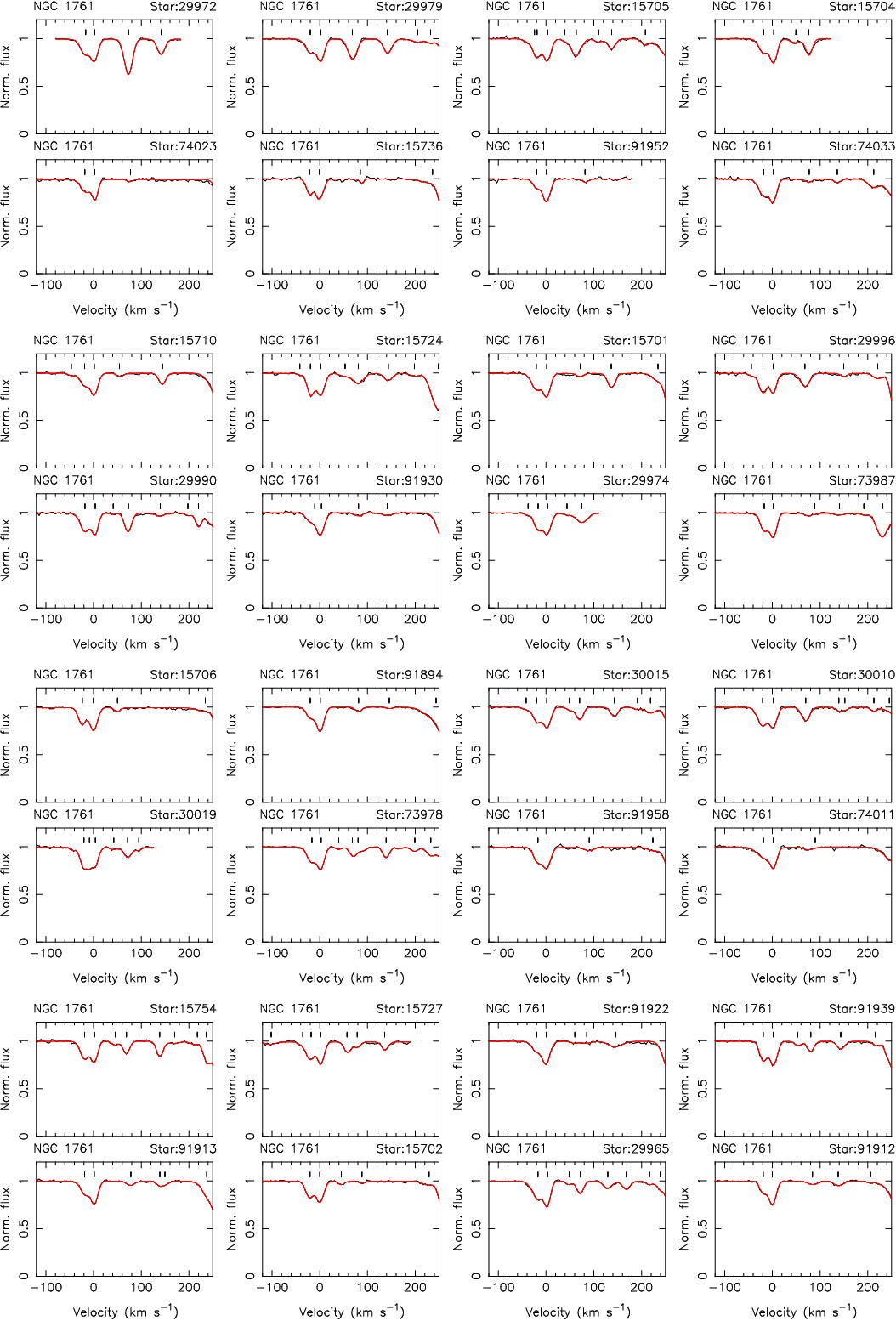}
\caption[]{FLAMES-GIRAFFE Ca\,{\sc ii} K spectra towards NGC\,1761 showing the spectra of 16 star-to-star 
pairs in which the maximum difference in the equivalent width was detected 
between a velocity of +30 and +105 km\,s$^{-1}$ in the LSR. Each star is only plotted once.}
\label{f_MaxVariation_EW_I/HVCs_NGC_1761_I/HVC1_p30_p105}
\end{figure*}

\begin{figure*}
\setcounter{figure}{3}
\includegraphics[height=22cm]{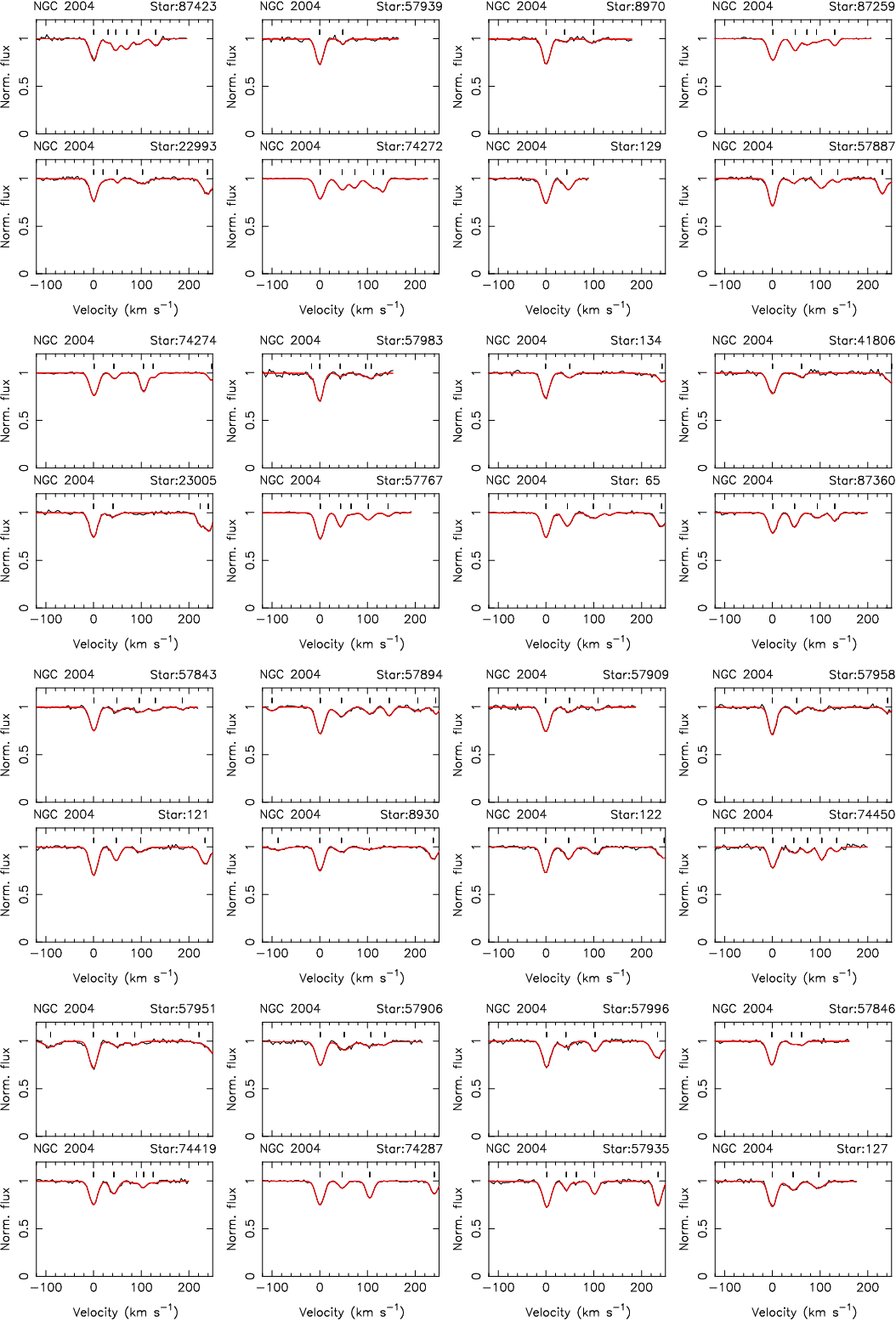}
\caption[]{FLAMES-GIRAFFE Ca\,{\sc ii} K spectra towards NGC\,2004 showing the spectra of 16 star-to-star 
pairs in which the maximum difference in the equivalent width was detected 
between a velocity of +30 and +75 km\,s$^{-1}$ in the LSR. Each star is only plotted once.}
\label{f_MaxVariation_EW_I/HVCs_NGC_2004_I/HVC1_p30_p75}
\end{figure*}

\begin{figure*}
\setcounter{figure}{4}
\includegraphics[height=22cm]{SpectraPlot_FLAMES_data_biggest_EW_Diff_NGC_2004_IHVC2_p75_p130.ps}
\caption[]{FLAMES-GIRAFFE Ca\,{\sc ii} K spectra towards NGC\,2004 showing the spectra of 16 star-to-star 
pairs in which the maximum difference in the equivalent width was detected 
between a velocity of +75 and +130 km\,s$^{-1}$ in the LSR. Each star is only plotted once.}
\label{f_MaxVariation_EW_I/HVCs_NGC_2004_I/HVC2_p75_p130}
\end{figure*}

\clearpage
\newpage

\begin{figure}
\setcounter{figure}{5}
\includegraphics[width=9cm]{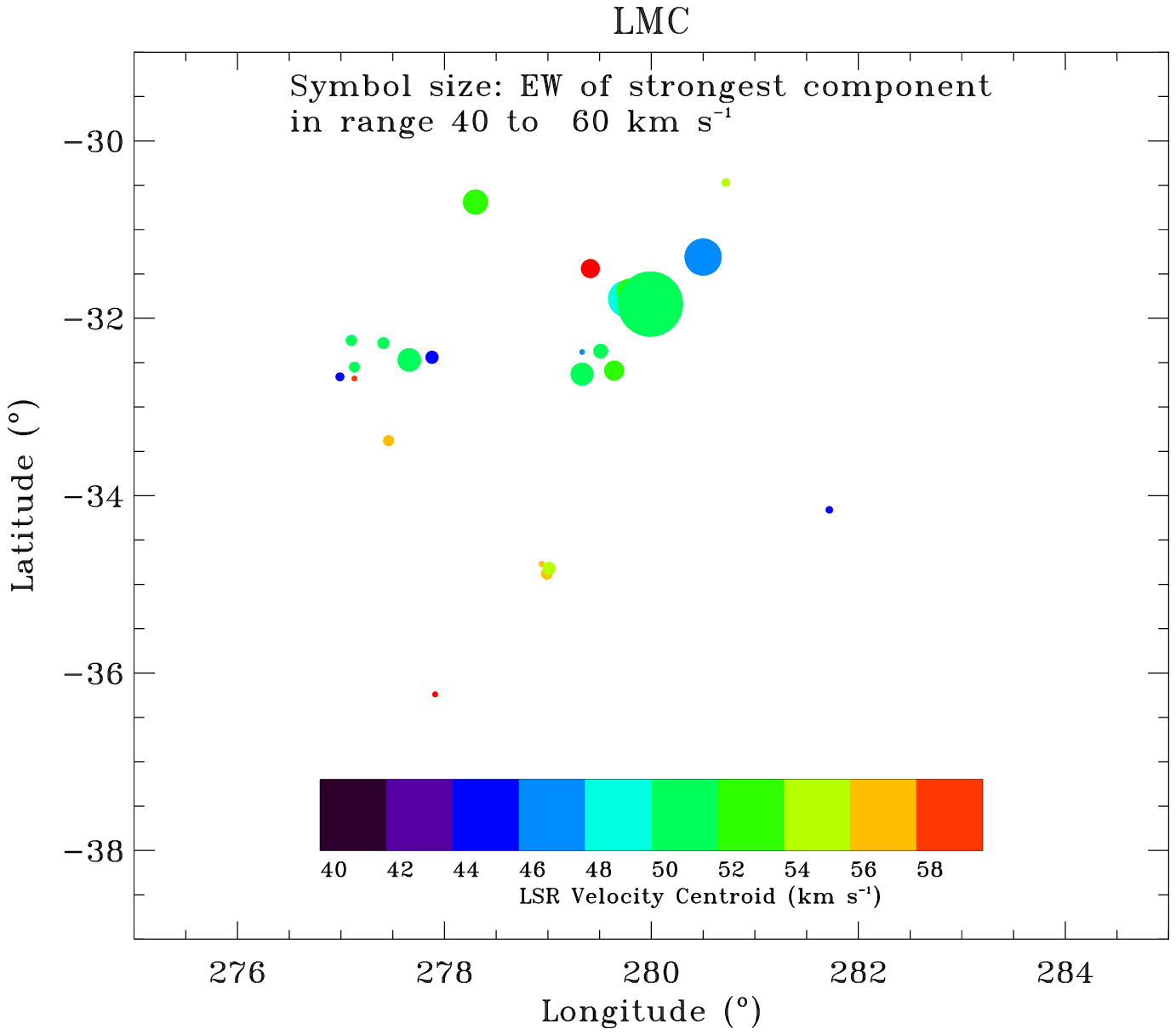}
\caption[]{Variation in the Ca\,{\sc ii} equivalent width for I/HVC components observed with FEROS 
towards the LMC for velocities between +40 and +60 km\,s$^{-1}$.} 
\label{f_logN_position_FEROS_1}
\end{figure}

\begin{figure}
\setcounter{figure}{6}
\includegraphics[width=9cm]{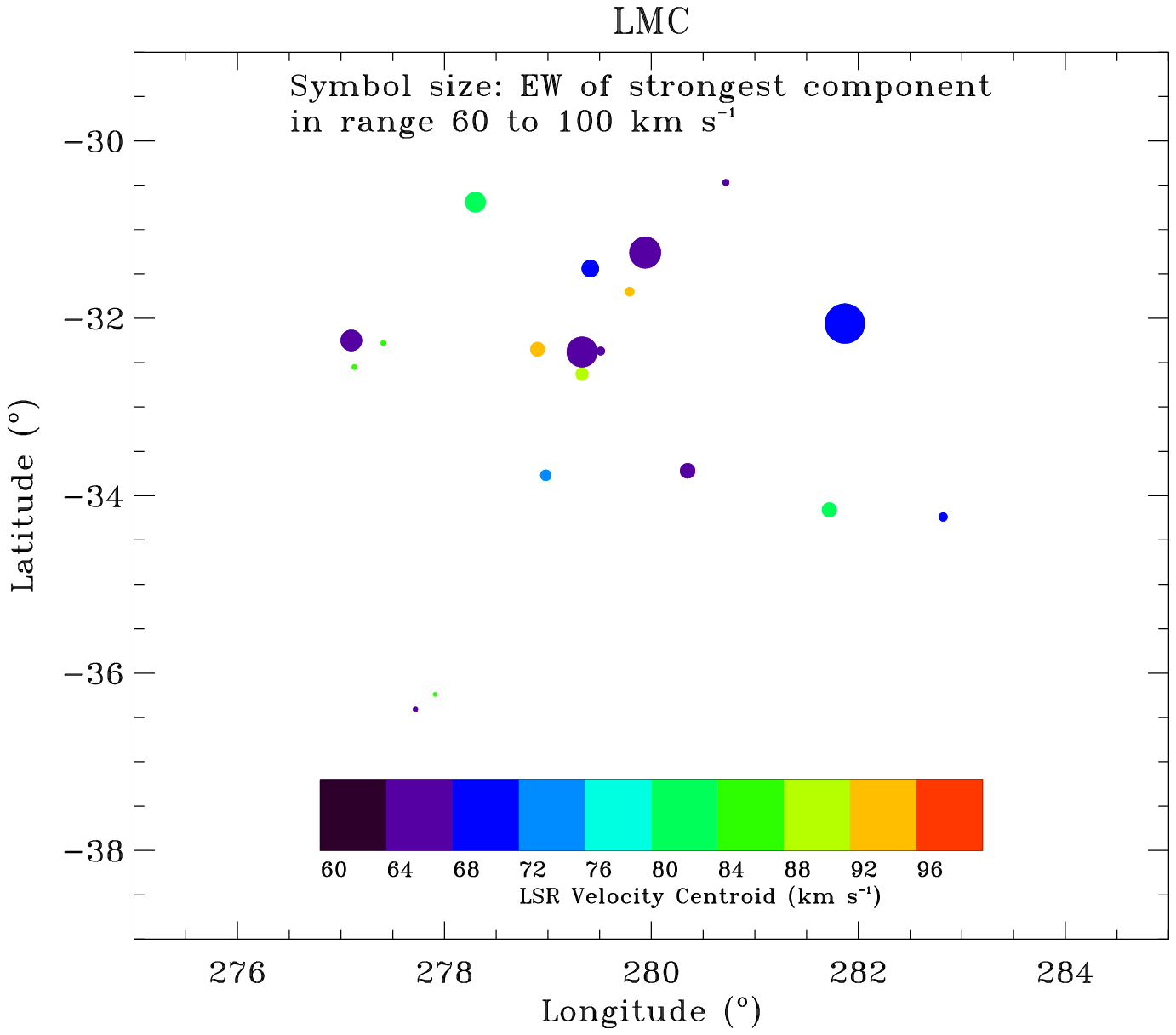}
\caption[]{Variation in the Ca\,{\sc ii} equivalent width for I/HVC components observed with FEROS 
towards the LMC for velocities between +60 and +100 km\,s$^{-1}$.} 
\label{f_logN_position_FEROS_2}
\end{figure}

\begin{figure}
\setcounter{figure}{7}
\includegraphics[width=9cm]{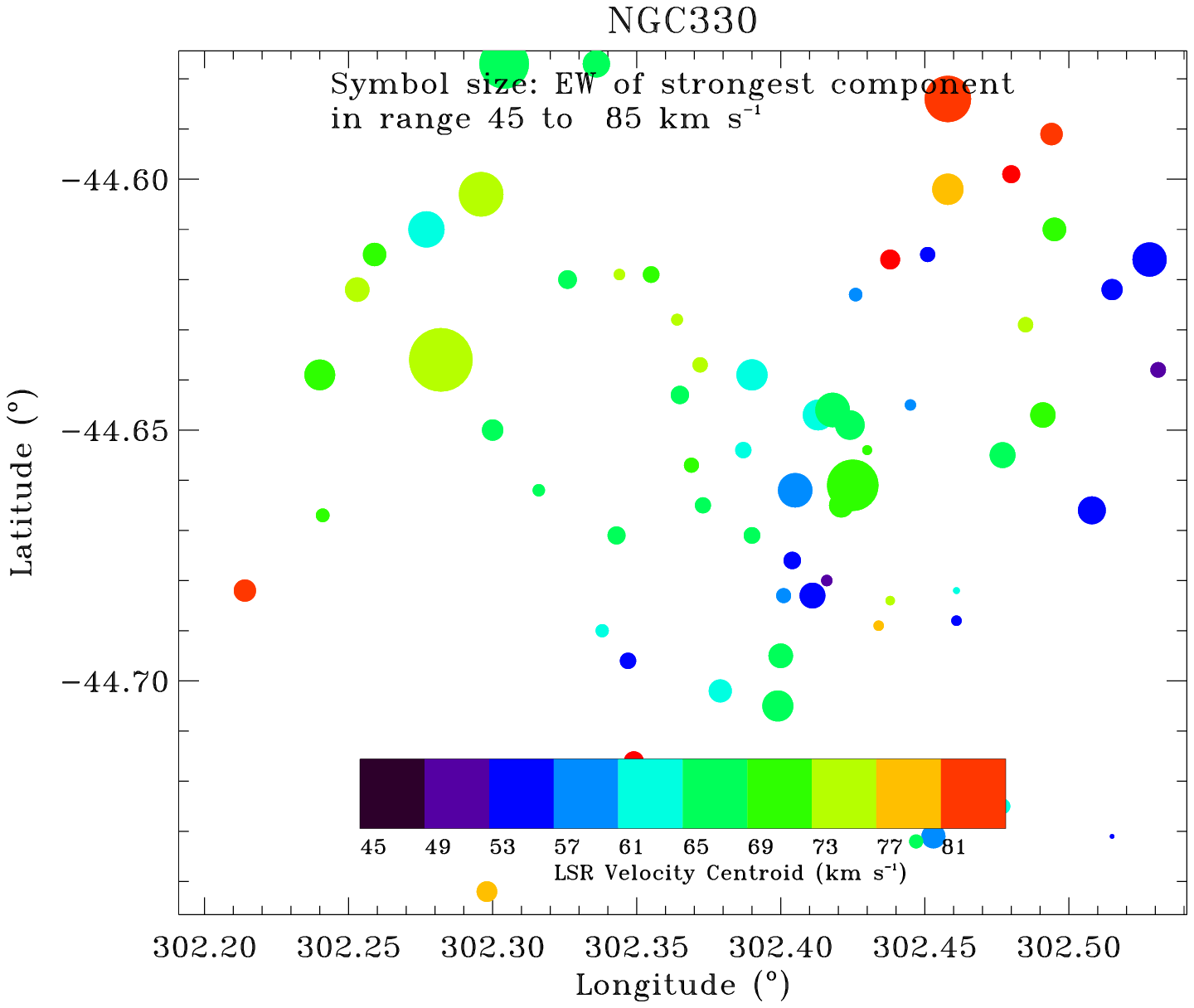}
\caption[]{NGC 330 equivalent width and peak velocity between +45 and +85 km\,s$^{-1}$.} 
\label{f_FLAMES_NGC_330_IVC_EW_Vel}
\end{figure}

\begin{figure}
\setcounter{figure}{8}
\includegraphics[width=9cm]{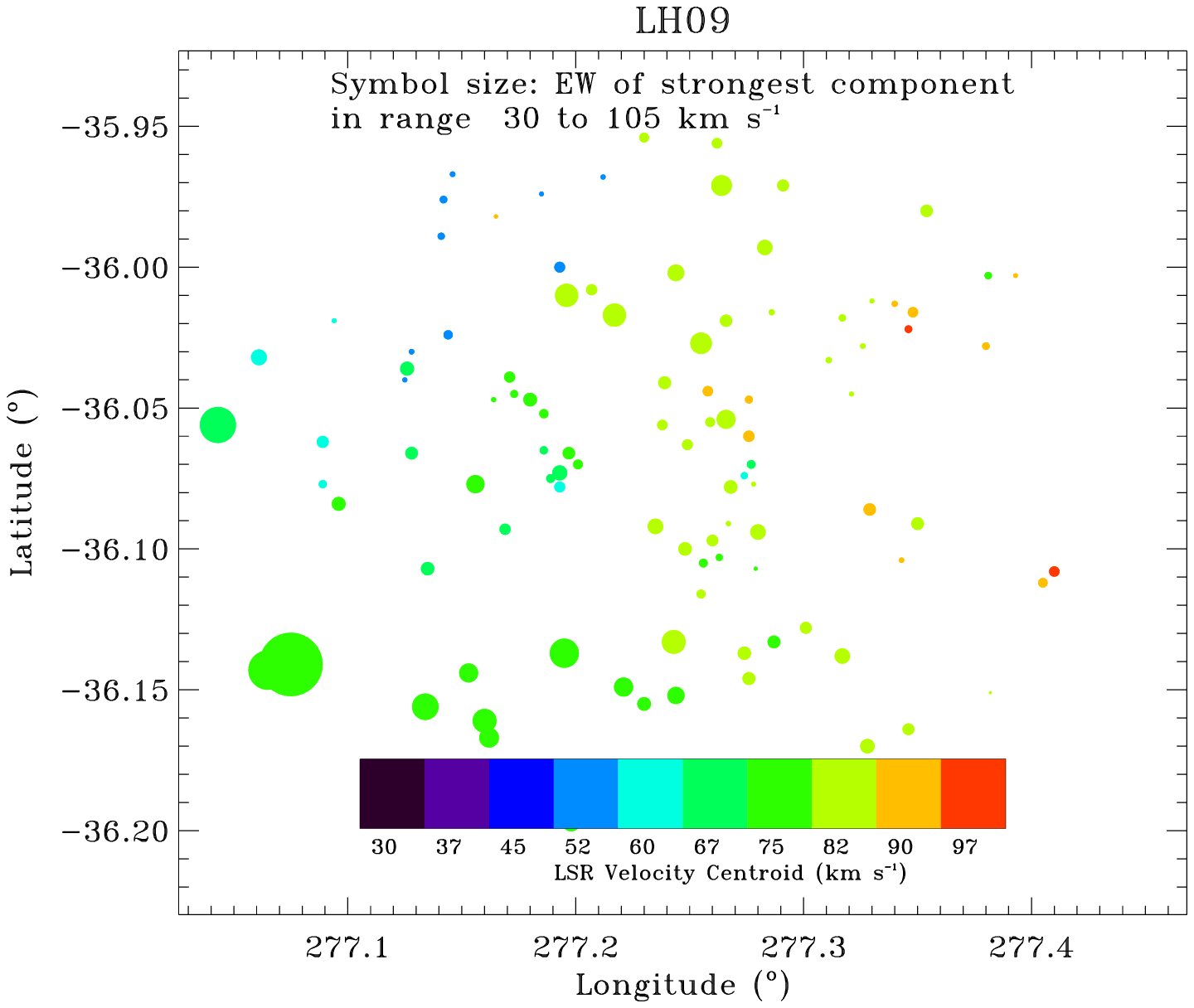}
\caption[]{NGC\,1761 equivalent width and peak velocity between +30 and +105 km\,s$^{-1}$.} 
\label{f_FLAMES_LH09_v_p30_p105_I_HVCs_EW_Vel}
\end{figure}

\begin{figure}
\setcounter{figure}{9}
\includegraphics[width=9cm]{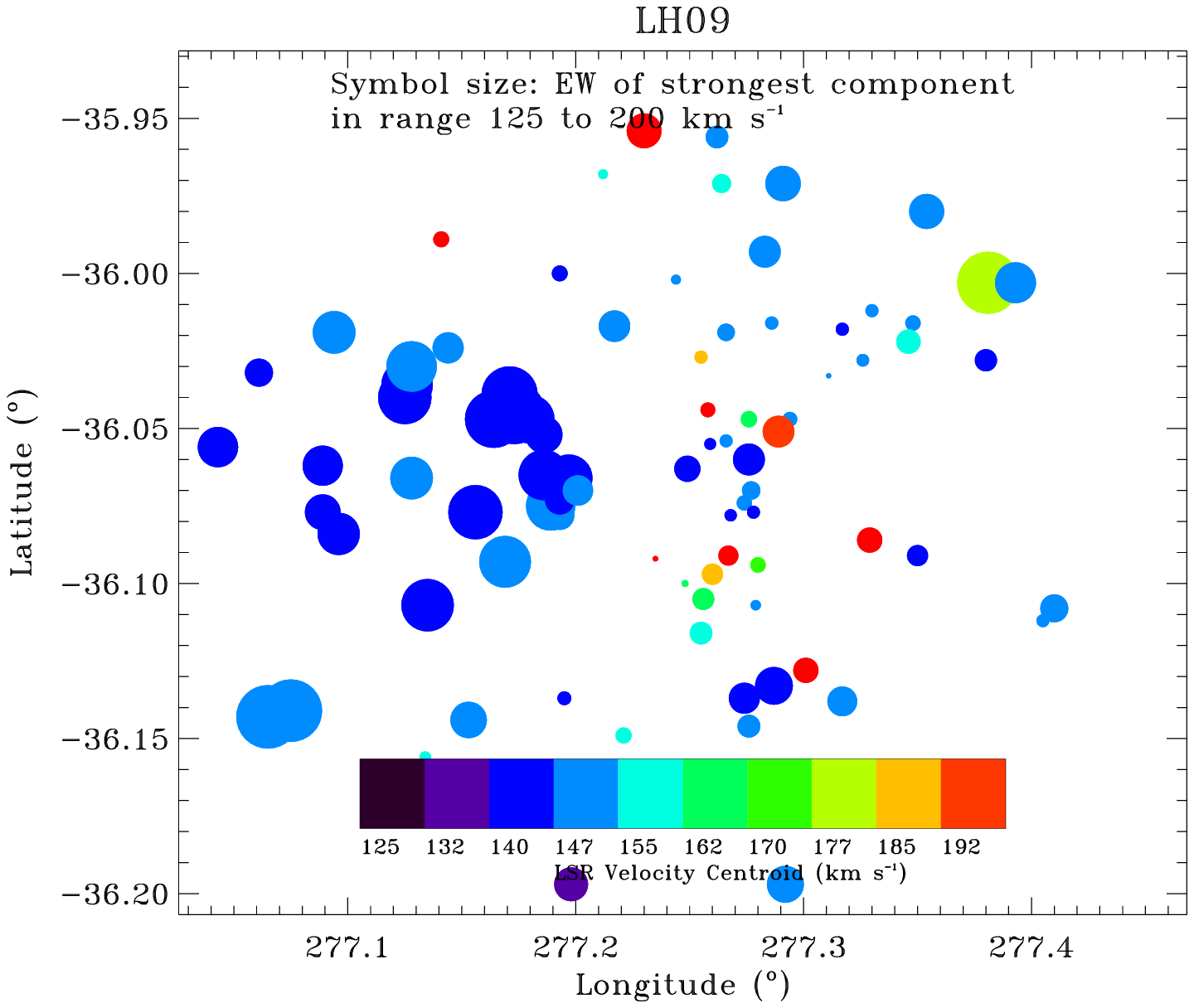}
\caption[]{NGC\,1761 equivalent width and peak velocity between +125 and +200 km\,s$^{-1}$.} 
\label{f_Fin_Prof_Coldns_Results_FLAMES_LH09_LMC1_p125_p200}
\end{figure}

\begin{figure}
\setcounter{figure}{10}
\includegraphics[width=9cm]{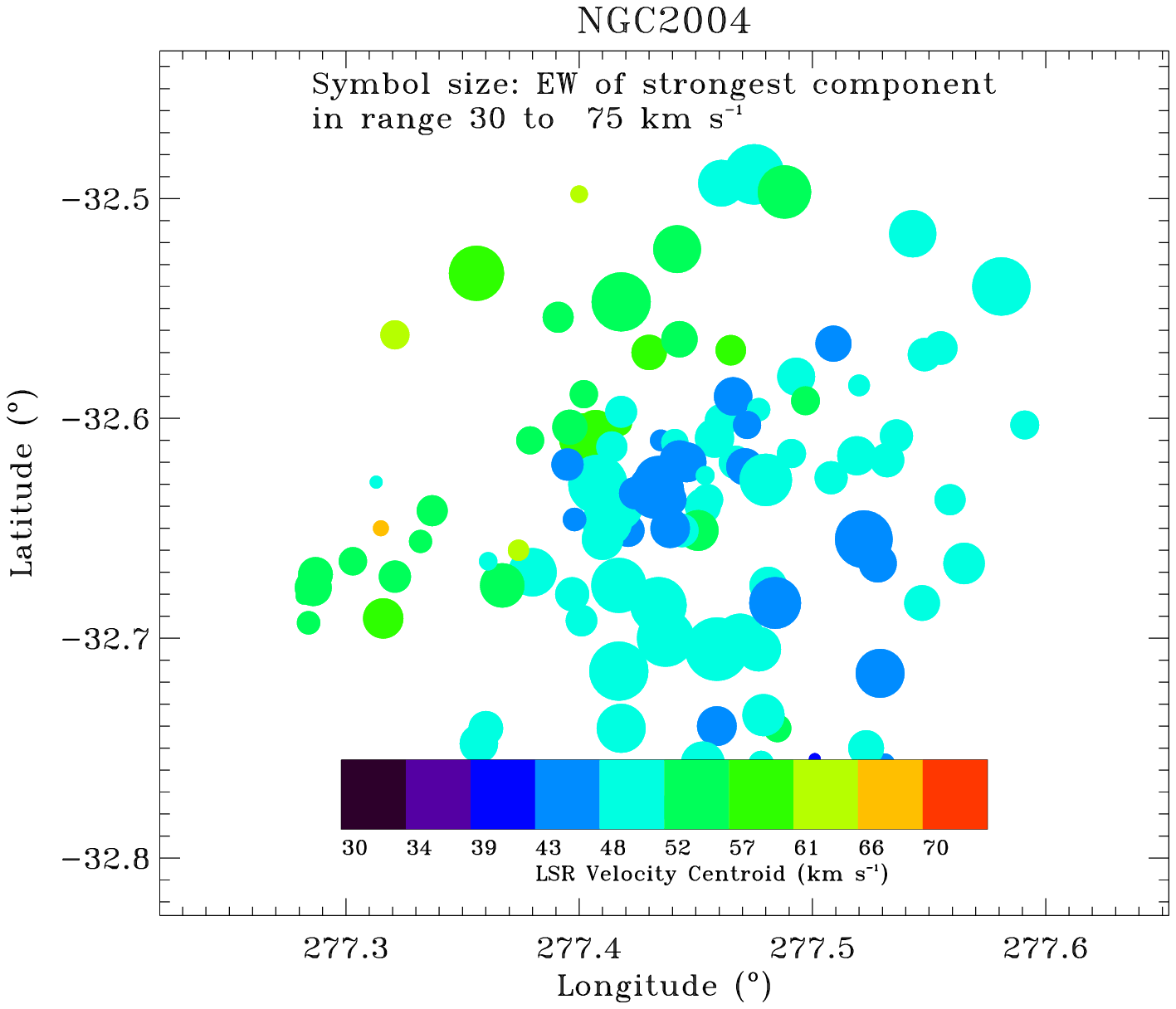}
\caption[]{NGC\,2004 equivalent width and peak velocity between +30 and +75 km\,s$^{-1}$.} 
\label{f_FLAMES_NGC2004_v_p30_p75_EW_Vel}
\end{figure}

\begin{figure}
\setcounter{figure}{11}
\includegraphics[width=9cm]{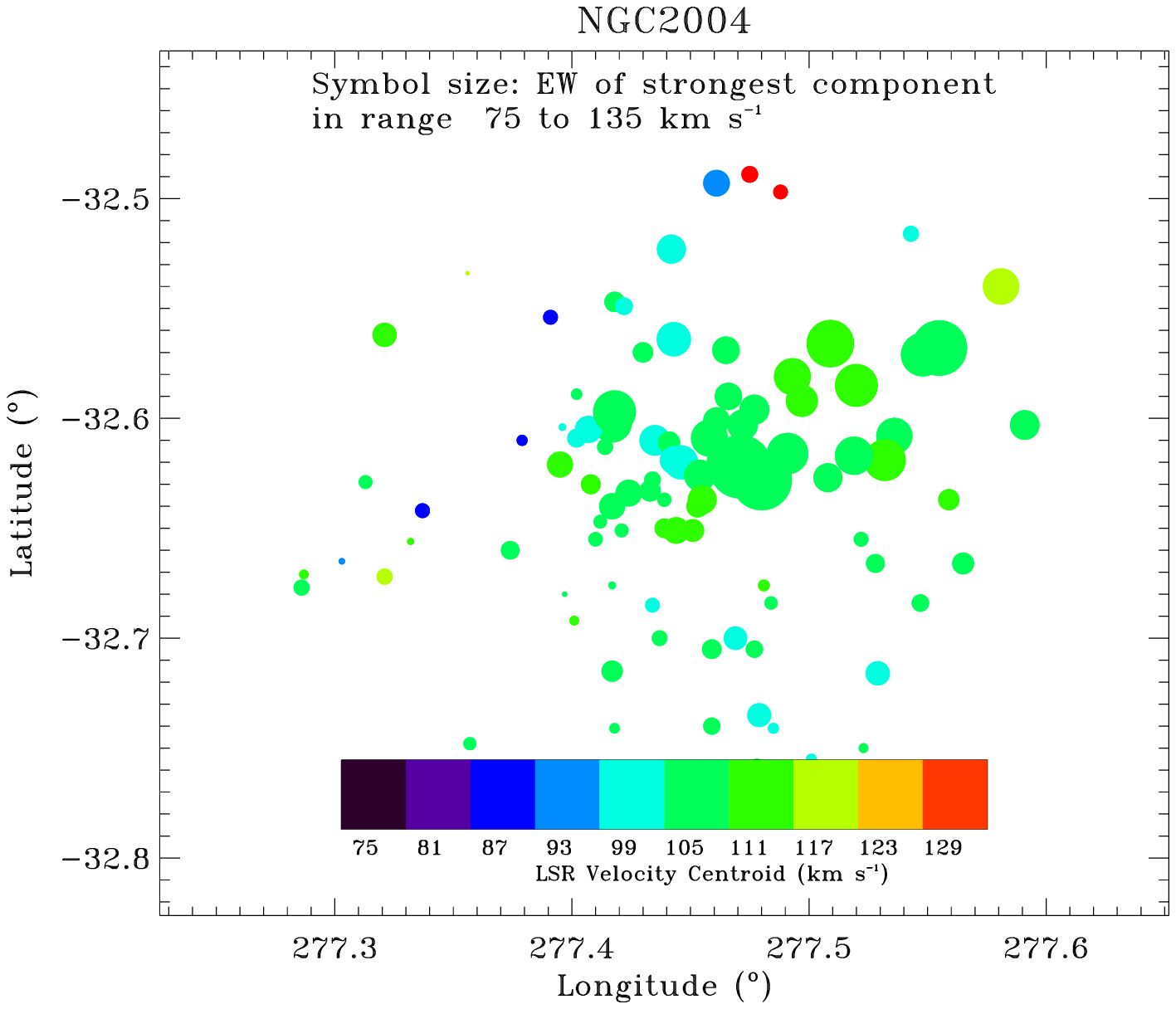}
\caption[]{NGC\,2004 equivalent width and peak velocity between +75 and +130 km\,s$^{-1}$.} 
\label{f_Fin_Prof_Coldns_Results_FLAMES_NGC2004_LMC1_p75_p130}
\end{figure}

\begin{figure}
\includegraphics[height=20cm]{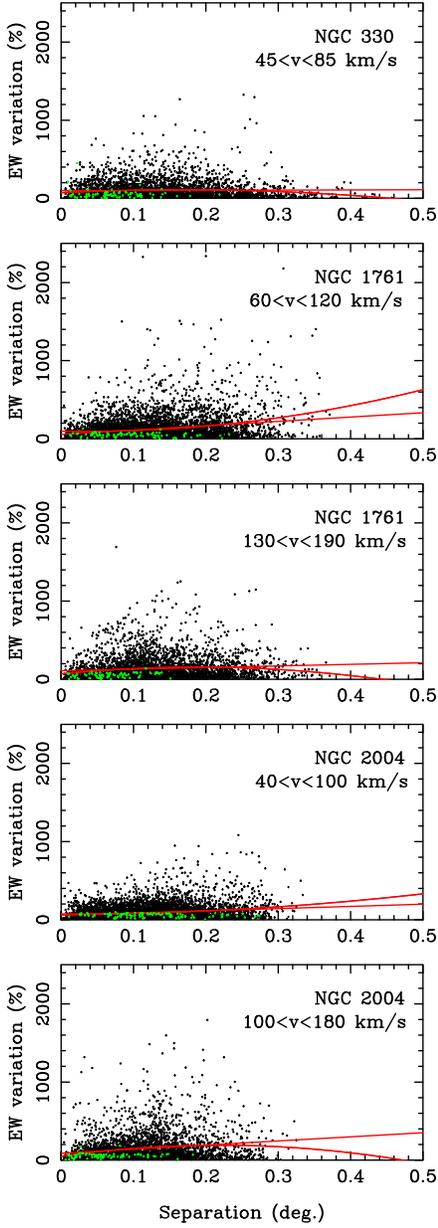}
\caption[]{Percentage equivalent width variation in Ca\,{\sc ii} (100.0$\times$(EW1/EW2)-100.0) plotted against star to star distance in degrees. 
(a) NGC 330: IVC from +45.0 to  +85.0 km\,s$^{-1}$ (b) NGC 1761: I/HVC from  +60.0 to +120.0 km (c) NGC 1761: HVC from +130.0 to +190.0 km/s 
(d) NGC 2004: I/HVC from  +40.0 to +100.0 km/s (e) NGC 2004: I/HVC from +100.0 to +180.0 km/s.}
\label{EWvariation}
\end{figure}


\begin{thebibliography}{}


%

\bibitem[2004]{and04} Andr\'{e} M.~K., et al., 2004, A\&A, 422, 483

%
%

\bibitem[1988]{adr88} Andreani P., Vidal-Madjar A., 1988, Nature, 333, 432

\bibitem[2009]{asp09} Asplund M., Grevesse N., Sauva, A.~J., Allende Prieto C., Kiselman D., 2004, 
A\&A, 47, 751

\bibitem[2003]{bag03} Bagnulo S., Jehin E., Ledoux C., Cabanac R., Melo C.,
Gilmozzi R., 2003, ESO Messenger no. 114, Page 10



\bibitem[2008]{ben08} Ben Bekhti N., Richter P., Westmeier T., Murphy, M.~T., 2008,
A\&A, 487, 583


\bibitem[2009]{ben09} Ben Bekhti N., Richter P., Winkel B., Kenn F., Westmeier T., 2009, 
A\&A, 503, 483


\bibitem[1992]{ber92} Bertelli G., Mateo M., Chiosi C., Bressan A., 1992,
ApJ, 388, 400



\bibitem[1980]{bla80} Blades J.~C., 1980, MNRAS, 190, 33


\bibitem[1988]{bla88a} Blades J.~C., Wheatley J.~M., Panagia N., Grewing M., Pettini M., Wamsteker W., 1988a, 
ApJ, 332, 75

%

\bibitem[1988]{bla88b} Blades J.~C., Wheatley J.~M., Panagia N., Grewing M., Pettini M., Wamsteker W., 1988b, 
ApJ, 334, 308

\bibitem[2009]{bla09} Bland-Hawthorn J., 2009, IAUS, 254, 241 

\bibitem[1999]{bli99} Blitz L., Spergel D.~N., Teuben P.~J., Hartmann D., Burton W.~B., 1999,
ApJ, 514, 818


\bibitem[2001]{blu01} Bluhm H., de Boer K.~S., Marggraf O., Richter P., 2001, A\&A, 367, 299


\bibitem[2004]{bre04} Bregman J.~N. 2004, in High Velocity Clouds, eds. H. van Woerden, B.~P. Wakker, 
U.~J. Schwarz, K.~S. de Boer, Astrophys. Space Sci. Lib., 312, 341



\bibitem[2008]{car08} Carrera R., Gallart C., Hardy E., Aparicio A., Zinn R., 2008, 
 AJ, 135, 836

\bibitem[2008]{chi08} Chiappini C., in Fumes J. G. S. J., Corsini E.M., Eds, ASP Conf. Ser. 2008, Vol. 396, 
Formation and Evolution of Galaxy Disks, Astron. Soc. Pac., San Francisco, p. 113



\bibitem[1992]{cra92} Crawford I.~A., 1992, MNRAS, 259, 47

\bibitem[2000]{dek00} Dekker H., D'Odorico S., Kaufer A., Delabre B., \& Kotzlowski H., 2000, SPIE, 4008, 534


\bibitem[2007]{des07} Dessauges-Zavadsky M., Combes F., Pfenniger D., 2007, A\&A, 473, 863

\bibitem[1989]{dia89} Diamond P.~J., Goss W.~M., Romney J.~D., Booth R.~S., Kalberla P.~M.~W., Mebold U., 1989, 
ApJ, 347, 302


\bibitem[2002]{deh02} de Heij V., Braun R., Burton W.~B., 2002, A\&A, 391, 67


\bibitem[2005]{fox05} Fox A.~J., Wakker B. P., Savage B. D., Sembach K. R., Tripp
T. M., \& Bland-Hawthorn J. 2005, ApJ, 630, 332

\bibitem[2010]{fox10} Fox A.~J., Wakker B.~P., Smoker J.~V., 
Richter P., Savage B.~D., Sembach K.~R., 2010, ApJ, 718, 1046

\bibitem[2013]{fox13} Fox A.~J., Richter P., Wakker B.~P., Lehner N., Howk C.~J., Ben Bekhti N., Bland Hawthorn J., Lucas S., 
2013, ApJ, 722, 110

\bibitem[2014]{fox14} Fox A.~J., Wakker B.~P., Barger K.~A., et al., 2014, ApJ, 787, 147


\bibitem[1981]{gio81} Giovanelli  R., 1981, AJ, 86, 1468

\bibitem[2014]{gri14} Gritton J.~A., Shelton R. L., Kwak K., 2014, 795, 99


\bibitem[2009]{hei09} Heitsch, F., Putman, M. E. 2009, ApJ, 698, 1485


\bibitem[1975]{hob75} Hobbs L.~M., 1975, ApJ, 202, 628


\bibitem[2007]{hop07} Hopp U., Schulte-Ladbeck R.~E., Kerp J., 2007, MNRAS, 374, 1164

\bibitem[2002]{how02} Howarth I.~D., Price R.~J., Crawford I.~A., Hawkins I., 2002,
MNRAS, 335, 267

\bibitem[2003]{how03} Howarth I.~D., Murray J., Mills D., Berry D.~S., 2003,
Starlink User Note SUN 50, Rutherford Appleton Laboratory/CCLRC

\bibitem[2006]{hun06} Hunter I., Smoker J.~V., Keenan F.~P., Ledoux C., Jehin E., Cabanac R., Melo C., Bagnulo S., 2006,
MNRAS, 367, 1478 

\bibitem[2005]{kal05} Kalberla P.~M.~W., Burton W.~B., Hartmann D., Arnal E.~M., Bajaja E., Morras R.,
P{\"o}ppel, W.~G.~L.\ 2005, 
A\&A, 440, 775


\bibitem[1999]{kau05} Kaufer A., Stahl O., Tubbesing S., Nørregaard P., Avila G., Francois P., Pasquini L., Pizzella A., 
1999, The Messenger 95, 8

\bibitem[2006]{kel06} Keller S.~C., Wood P.~R., 2006, ApJ, 642, 834

\bibitem[1998]{ken98} Kennicut R.~C., 1998, ApJ, 498, 541

\bibitem[2007]{lau07} Lauroesch J. T., 2007, in Haverkorn M., Goss W. M., eds, ASP Conf. Ser.
Vol. 365, Small Ionized and Neutral Structures in the Diffuse Interstellar Medium, Astron. Soc. Pac., 
San Francisco, p. 40


\bibitem[2009]{leh09} Lehner N., Staveley-Smith L., Howk J.~C., 2009, ApJ, 702, 940

\bibitem[2011]{leh11} Lehner N., Howk J.~C., 2011, Science, 334, 955





\bibitem[2011]{mat11} Marasco A., Fraternali F., 2011, A\&A, 525, 134

%


\bibitem[2009]{mcc09} McClure-Griffiths N.~M., et al., 2009, ApJS, 181, 398

\bibitem[1999]{mey09} Meyer D.~M., Lauroesch J.~T., 1999, 520, 103

\bibitem[1981]{mir81}  Mirabel I.~F., 1981, ApJ, 250, 528

\bibitem[2009]{mis09} Misawa T., Charlton J.~C., Kobulnicky H.~ A., Wakker, B.~P., Bland-Hawthorn J., 2009, 
ApJ, 695, 1382

\bibitem[mol93]{mol93} Molaro P., Vladilo G., Monai S., d'Odorico S., Ferlet R., Vidal-Madjar A., Dennefeld M., 1993, A\&A, 274, 505

\bibitem[2003]{mor03} Morton D.~C., 2003, ApJS, 149, 205

\bibitem[2004]{mor04} Morton D.~C., 2004, ApJS, 151, 403

\bibitem[1963]{mul63} Muller C.~A., Oort J.~H., Raimond E., 1963, 
CR Acad. Sci. Paris, 257, 1661


\bibitem[2010]{nas10} Nasoudi-Shoar S., Richter P., de Boer K.~S., Wakker B.~P., 2010, 
A\&A, 520, 26


\bibitem[2008]{nid08} Nidever D.~L., Majewski S.~R., Burton W.~B., 2008, ApJ, 679, 432


\bibitem[2004]{ola04} Olano C.~A., 2004, A\&A, 423, 895

%
%

\bibitem[2008]{olo08} Olano C.~A., 2008, A\&A, 485, 457

%
%
\bibitem[2002]{pas02} Pasquini, L. et al. 2002, The Messenger 110, 1


%
%

\bibitem[2004]{pis04} Pisano D.~J., Barnes D.~G., Gibson B.~K., Staveley-Smith L., Freeman K.~C., Kilborn V.~A., 
2004, ApJ, 610, L17



\bibitem[2004]{poi04} Points S.~D., Lauroesch J.~T., Meyer D.~M., 2004, PASP, 116, 801

%
%

\bibitem[2010]{ri10} Richings A.~J., Schaye J, Oppenheimer B.~D., 2010, MNRAS, 440, 3349

%
%
%
\bibitem[1999]{ric99} Richter P., de Boer K.~S., Widmann H., Kappelmann N., Gringel W., Grewing M., Barnstedt J., 1999, 
Nature, 402, 386

%
%
\bibitem[2003]{ric03} Richter P., Sembach K.~R., Howk J.~C., 2003, A\&A, 405, 1013

\bibitem[2009]{ric09} Richter P., Charlton J.~C., Fangano A.~P.~M., Bekhti N.~B., Masiero J.~R., 2009, 
ApJ, 695, 1631

\bibitem[2011]{ric11} Richter P., Krause F., Fechner C., Charlton J.~C., Murphy M.~T., 2011, A\&A, 528, 12



\bibitem[2014]{ric14} Richter P., Fox A.~J., Ben Bekhti N., Murphy M.~T., Bomans D., Frank S., AN, 335, 92

\bibitem[1952]{rs52} Routly P.~M., Spitzer L., 1952,
ApJ, 115, 227

%
%
\bibitem[1981]{sav81} Savage B.~D., de Boer K.~S., 1981, ApJ, 243, 460

\bibitem[2000]{sem00} Sembach K.~R., Howk C.~J., Ryans R.~S.~I., Keenan F.~P., 2000, 
ApJ, 528, 310


\bibitem[2005]{sie05} Siegel M.~H., Majewski S.~R., Gallart C., Sohn S.~T., Kunkel W.~E., Bran R., 2005, ApJ, 623, 181

\bibitem[1974]{sil74} Siluk R. S., Silk J., 1974,
ApJ, 192, 51


\bibitem[2002]{sim02} Simon J.~D., Blitz L., 2002, ApJ, 574, 726

\bibitem[2001]{smo01} Smoker J.~V., Roger R.~S., Keenan F.~P., Davies R.~D., Lang R.~H., Bates B., 2001, 
A\&A, 380, 683



\bibitem[2003]{smo03} Smoker J.~V., et al., 2003, MNRAS, 346, 119


%
%

\bibitem[2011]{smo11b} Smoker J.~V., Bagnulo S., Cabanac R., Ledoux C., Jehin E., Melo S., Keenan F.~P., 2011, 
MNRAS, 414, 59

\bibitem[2014]{smo14} Smoker J.~V., Ledoux, C. Jehin E. Keenan F.~P., Kennedy M., Cabanac R., Melo, C., 2014, 
MNRAS, 438, 1127

\bibitem[2015]{smo15} Smoker J.~V., Keenan F.~P., Fox A., 2015, 582, 59


\bibitem[1981]{son81a} Songaila A., 1981, ApJ, 243, L19


\bibitem[1981]{son81b} Songaila A., Cowie L.~L., York D.~G., 1981, ApJ, 248, 956


\bibitem[1986]{son86} Songaila A., Blades J.~C., Hu E.~M., Cowie L.~L., 1986, 
ApJ, 303, 198

\bibitem[2008]{sta08} Stanimirovi\'c S., Hoffman S., Heiles C., Douglas K.~A., Putman, M., Peek J.~E.~G., 2008, 
ApJ, 680, 276


\bibitem[2003]{sta03} Staveley-Smith L., Kim S., Calabretta M.~R., Haynes R.~F., Kesteven M., 2003, 
MNRAS, 339, 87


\bibitem[2006]{tom06} Thom C., Putman M.~E., Gibson B.~K., Christlieb N., Flynn C., Beers T.~C., 
Wilhelm R., Lee Y.~S., 2006, 
ApJ, 638, L97

\bibitem[2008]{tom08} Thom C., Peek J.~E.~G., Putman M.~E., Heiles Carl., Peek K.~M.~G., Wilhelm R., 2008,
ApJ, 684, 364

\bibitem[1993]{val93} Vallerga J.~V., Vedder P.~W., Craig N., Welsh B.~Y., 1993,
ApJ, 411, 729


\bibitem[1999]{van99} van Loon J.~Th., Smith K.~T., McDonald I., Sarre P.~J., Fossey S.~J., Sharp R.~G., 2009, MNRAS, 399, 195

\bibitem[2013]{van13} van Loon J.~Th., Bailey M., Tatton B.~L., et al., 2013, A\&A, 550, 108

\bibitem[1999]{van99} van Woerden H., Schwarz U.~J., Peletier R.~F., Wakker B.~P., Kalberla P.~M.~W., 1999,
Nat., 400, 138


\bibitem[1997]{wak97} Wakker B.~P., van Woerden H., 1997, ARA\&A, 35, 217


\bibitem[2000]{wak00} Wakker B.~P., Mathis J.~S., 2000, 
ApJ, 544, 107

\bibitem[2001]{wak01a} Wakker B.~P., 2001, ApJS, 136, 463



\bibitem[2007]{wak07} Wakker B.~P., et al., 2007, ApJ, 670, 113

\bibitem[2008]{wak08} Wakker B.~P., York D.~G., Wilhelm R., Barentine J.~C., Richter P., Beers T.~C., 
 Ivezic Z., Howk J.~C., 2008, ApJ, 672, 298

\bibitem[1996]{wal96} Wallace P., Clayton C., 1996, {\sc rv}, Starlink User Note
SUN 78, Rutherford Appleton Laboratory/CCLRC




\bibitem[1990]{way90} Wayte S.~R., 1990, ApJ, 355, 473


\bibitem[2000]{weg00} Wegner et al., 2000, A\&AS, 143, 9

\bibitem[1999]{wel99} Welsh B.~Y., Wheatley J., Lallement R., 2009, PASP, 121, 606


\bibitem[1997]{wel97} Welty D.~E., Lauroesch J.~T., Blades J.~C., Hobbs L.~M., York D.~G.,
1997, ApJ, 489, 672

%
%
\bibitem[1999]{wel09} Welty D.~E., Frisch P.~C., Sonneborn G., York D.~G., 1999, 
ApJ, 512, 636


\bibitem[2001]{wel01} Welty D., Fitzpatrick E.~L., 2001, ApJ, 551, 175

%

\bibitem[2006]{wel06} Welty D.~E., Federman S.~R., Gredel R., Thorburn J.~A., Lambert D.~L., 2006, 
ApJS, 165, 138 


\bibitem[2012]{wel12} Welty D.~E., Xue R., Wong T., 2012, ApJ, 745, 173


\bibitem[2008]{wes08} Westmeier T., Br\"uns C., Kerp J., 2008, MNRAS, 390, 1691

\end{thebibliography}
\end{document}